\documentclass[a4paper, 11pt]{article}
\usepackage{graphicx} 
\usepackage[english]{babel}
\usepackage[toc,page]{appendix}
\usepackage{color}
\usepackage[utf8]{inputenc}
\usepackage[T1]{fontenc}
\usepackage{amsmath, amsthm, amssymb}
\usepackage{mathtools}
\usepackage{mathrsfs}
\usepackage{bbm}
\usepackage{enumitem}
\usepackage{hyperref}
\usepackage{geometry}
\usepackage{dsfont}

\usepackage[dvipsnames]{xcolor}

\usepackage{thmtools}
\usepackage[normalem]{ulem}

\usepackage{import}
\usepackage[doi=true,isbn=false, url=true, date=year, backend=biber,maxbibnames=99
]{biblatex} 
\renewbibmacro*{journal+issuetitle}{%
  \usebibmacro{journal}%
  \setunit*{\addspace}%
  \iffieldundef{series}
    {}
    {\newunit
     \printfield{series}%
     \setunit{\addspace}}%
  \usebibmacro{volume+number+eid}%
  \setunit{\bibpagespunct}%
  \printfield{pages}%
  \setunit{\addspace}%
  \usebibmacro{issue+date}%
  \setunit{\addcolon\space}%
  \usebibmacro{issue}%
  \newunit}

\renewbibmacro*{note+pages}{%
  \printfield{note}%
  \newunit}

\theoremstyle{plain}
\newtheorem{theorem}{Theorem}[section]
\newtheorem{lemma}[theorem]{Lemma}
\newtheorem{proposition}[theorem]{Proposition}

\theoremstyle{definition}
\newtheorem{definition}[theorem]{Definition}

\theoremstyle{remark}
\newtheorem{remark}[theorem]{Remark}

\DeclareMathOperator{\aut}{Aut}

\newcommand{\E}{\mathbb{E}}
\newcommand{\tr}{\mathrm{tr}}
\newcommand{\e}{\mathrm{e}}
\newcommand{\diam}{\mathrm{diam}}
\newcommand{\dd}{\mathrm{d}}
\renewcommand{\d}{\mathrm{d}}
\renewcommand{\i}{\mathrm{i}}
\renewcommand{\Phi}{\varPhi}
\renewcommand{\Psi}{\varPsi}
\renewcommand{\Sigma}{\varSigma}
\newcommand{\epsi}{\varepsilon}

\newcommand{\mI}{\mathcal{I}}

\newcommand{\w}{\omega}
\newcommand{\N}{\mathbb{N}}
\newcommand{\Z}{\mathbb{Z}}
\newcommand{\C}{\mathbb{C}}

\newcommand{\mA}{\mathcal{A}}
\newcommand{\mU}{\mathfrak{U}}
\newcommand{\mV}{\mathcal{V}}
\newcommand{\R}{\mathbb{R}}
\newcommand{\eps}{\varepsilon}

\newcommand{\OD}[1][]{   ^{\mathrm{OD}#1}   }

\newcommand{\mL}{\mathcal{L}}
\newcommand{\ad}{\mathrm{ad}}

\usepackage{mathtools}

\DeclarePairedDelimiter\norm{\lVert}{\rVert}

\DeclarePairedDelimiter\br{\lparen}{\rparen}

\newcommand\numberthis{\addtocounter{equation}{1}\tag{\theequation}}

\usepackage{parskip}

\title{The generalized adiabatic theorem for extended lattice systems}
\author{Lennart Becker%
\texorpdfstring{\footnote{\parbox[t]{.7\textwidth}{
                \foreignlanguage{ngerman}{Fachbereich Mathematik,  Universität Tübingen,\\
                Auf~der~Morgenstelle~10, 72076~Tübingen,} Germany
            }
        }
    }{}%
\and Stefan Teufel%
\texorpdfstring{%
        \footnotemark[1]
    }{}%
\and Marius Wesle%
\texorpdfstring{%
        \footnotemark[1]
    }{}%
}
\date{\today}

\begin{document}

\maketitle

\begin{abstract}
We prove an adiabatic theorem for infinitely extended lattice fermion systems with gapped ground states, allowing perturbations that may close the gap. The Heisenberg dynamics on the CAR-algebra   is generated by a time dependent two-parameter family of Hamiltonians
$
H^{\varepsilon,\eta}_t=\eta^{-1}\!\bigl(H_t+\varepsilon(H^1_t+V_t)\bigr)
$, where $H_t$ is assumed to have a gapped ground state $\w_t$, $\eta\in(0,1]$ is the adiabatic parameter and $\eps\in[0,1]$ controls the strength of the perturbation.
We construct a quasi-local dressing transformation
$
\beta^{\varepsilon,\eta}_t=\exp \bigl(i \mathcal{L}_{S^{\varepsilon,\eta}_t}\bigr)
$
that yields super-adiabatic states 
\(\omega^{\varepsilon,\eta}_t
=\omega_t\circ\beta^{\varepsilon,\eta}_t\) 
which, when tested against local observables,  solve the corresponding time-dependent Schr\"odinger equation up to errors asymptotically smaller than any power of $\eta$ and $\eps$.
 The construction is local in space and time, does not assume uniqueness of the ground state, and works under super-polynomial decay of the interactions $H_t$ and $H_t^1$   rather than exponential decay. If the Hamiltonian is time-independent on an interval, the dressed state is \(\eta\)-independent and forms a non-equilibrium almost-stationary state with lifetime of order \(\varepsilon^{-\infty}\). 
 The result provides a rigorous basis for linear response to macroscopic changes in gapped systems, including a proof of  Ohm's law for macroscopic Hall currents.
\end{abstract}

\section{Introduction}

We prove a generalized adiabatic theorem for infinitely extended fermionic systems on the lattice~$\Z^d$ with a gapped ground state. We additionally allow for perturbations that close the spectral gap, meaning that our result is actually an adiabatic theorem for many-body resonances, called non-equilibrium almost-stationary states (NEASS) in this context.  Our result generalizes recent  works \cite{bachmann2017adiabatic,BDF18,T2020,henheik2022adiabatic,Henheik_Teufel_2022} in several ways.
In particular, we do not assume uniqueness of the ground state, allow for interactions that decay faster than any polynomial instead of requiring exponential decay, and keep track explicitly of the localization of all perturbations.
 We also provide a streamlined proof that does not rely on finite-volume approximations. 
Two new technical ingredients to the proof are recent results on the automorphic equivalence of gapped ground states in infinitely extended fermionic systems \cite{Becker_Teufel_Wesle_2025}, and Lieb–Robinson bounds for systems of fermions with long-range interactions \cite{TeufelWessel25}. The most important application of our adiabatic theorem is presumably the determination and justification of formulas for linear response coefficients in such systems. In particular, it is used to demonstrate Ohm's law for macroscopic Hall currents in \cite{wmmmt2024}.

Let us describe the setup and our result in some more detail. 
We consider a time-dependent Hamiltonian 
$I\ni t\mapsto H_t$ that acts as a derivation on the fermionic CAR-algebra $\mA$ over the lattice $\Z^d$. The two key assumptions are that $H_t$ is a $B_\infty$-interaction, i.e.\ the local terms of $H_t$ decay faster than any inverse polynomial in the diameter of their support,  and that there exists a differentiable family $I\ni t\mapsto \w_t$ of gapped ground states for $H_t$. Here $I\subseteq \R$ is an interval. In addition, we allow for time-dependent perturbations  $I\ni t\mapsto H^1_t$ and $I\ni t\mapsto V_t$, where $H^1_t$ is a second $B_\infty$-interaction and $V_t$ an external Lipschitz potential, cf.\ Definition~\ref{def:lipschitz potential}. We are interested in the Heisenberg dynamics generated by the time-dependent Hamiltonian
\[
H^{\varepsilon, \eta}_t = \tfrac{1}{\eta}(H_t  + \varepsilon(H^1_t + V_t))\,,
\]
i.e.\ in the cocycle $(\mU_{s, t}^{\varepsilon, \eta})_{(s,t)\in I^2}$ of automorphisms of $\mA$ generated by the time-dependent derivation $\i\mL_{H^{\varepsilon, \eta}_t}$, in the adiabatic regime $\eta\ll1$ and for small perturbations $\eps\ll 1$. Note that it is not assumed that $H^{\varepsilon, \eta}_t$ has a gapped ground state, i.e.\ the perturbations may close the spectral gap of $H_t$.

Our main result is the construction of a $B_\infty$-interaction $I\ni t\mapsto S^{\eps,\eta}_t$ that depends locally in time and space on $\dot H_t$, $H^1_t$, and $V_t$ and generates a family $I\ni t\mapsto \beta^{\eps,\eta}_t := \exp(\i \mL_{S^{\eps,\eta}_t})$ of near identity automorphisms of $\mA$ such that the dressed state $\w_t^{\eps,\eta} := \w_t \circ \beta^{\eps,\eta}_t$ is an approximate solution of the corresponding Schr\"odinger evolution whenever $\eps$ and $\eta$ are small. More precisely, it holds  for all strictly local observables $A\in\mA_0$, all $\alpha>0$, and all times $t,t_0\in I$ that
\[
\sup_{\eps\in [0,\eta^\alpha]}\left|\w_t^{\varepsilon, \eta}(A) - \w_{t_0}^{\varepsilon, \eta} (\mU_{t_0, t}^{\varepsilon, \eta} \, A) \right| = \mathcal{O}(\eta^\infty)\,.
\]
See Theorem~\ref{Theorem: main theorem} for a more explicit and quantitative bound. From the latter it also follows that, if for some time-interval $J\subseteq I$ the Hamiltonian $H^{\epsi,\eta}_t\equiv H^{\eps,\eta}_J$
is constant, then the dressed state $\w_t^{\eps,\eta}\equiv\w_J^{\eps,0} $ is constant as well and independent of $\eta$. Moreover, $\w_J^{\eps,0}$ is a non-equilibrium almost-stationary state (NEASS) for $H^{\eps,1}_J$ with life-time of order $1/\eps^\infty$. More precisely, in this situation there is an   exponent $m\leq 4d+2$ such that for all $t,t_0\in J$ and  $A\in\mA_0$  
\[
         \left|\w_J^{\varepsilon, 0}(A) - \w_J^{\varepsilon, 0} (\mU_{t_0, t}^{\varepsilon, 1} \, A) \right| = \mathcal{O}(\epsi^\infty \,|t-t_0|^m)\,.
        \]
Thus, even when the spectral gap closes due to the perturbation $\eps(H_t^1 + V_t)$, the system adiabatically evolves into an almost stationary state $\w_J^{\eps,0}= \w_J\circ \beta_J^{\eps,0}$ that is a small local deformation of the gapped ground state $\w_J$ of $H_J$.
Indeed, one main application of our result is the response of systems with gapped ground states to adiabatically switching on a potentially global perturbation that may close the spectral gap. In \cite{wmmmt2024,TW2025} Theorem~\ref{Theorem: main theorem} is used as a starting point to prove Ohm's law for macroscopic resp.\ microscopic Hall currents in infinite-volume models of interacting lattice fermions.

We  briefly comment on the extensive mathematical literature on adiabatic theorems. Due to the vast amount of material, we will not attempt to provide an exhaustive list of references. The standard adiabatic theorem for gapped systems was first proved in \cite{Kato} and systematic expansions to all orders were established, e.g., in \cite{Nenciu,JP}. Space adiabatic theorems, i.e.\ versions allowing for perturbations similar to Lipschitz potentials, were established, e.g., in \cite{EW,NS2004,PST}. Adiabatic theorems for quantum resonances, i.e.\ finite system versions of adiabatic theorems for NEASS, have been shown, e.g., in \cite{AF2007,EH2011}. For more details on the history of adiabatic theorems in quantum mechanics, see, for example, \cite{Teufel2022}.
Until recently, all adiabatic theorems were concerned with norm approximations of solutions to the Schr\"odinger equation, for which the corresponding error bounds are not uniform across different system sizes in many-body quantum systems.
The first adiabatic theorem for extended systems with uniform error control over the system size was formulated and proven in \cite{bachmann2017adiabatic} for finite quantum spin systems.  In \cite{Monaco_Teufel_2019}   a super-adiabatic version of \cite{bachmann2017adiabatic} was established for lattice fermions. In both works $\eps=0$ was assumed. The situation involving additional perturbations, $H^1_t$ and $V_t$, which could potentially close the spectral gap of $H_t$, was considered   first   in \cite{T2020}, where the concept of NEASS was introduced. Finally, in \cite{Henheik_Teufel_2022,henheik2022adiabatic} the result of \cite{T2020} was lifted to infinitely extended systems, however, assuming either a uniform gap also in finite volume \cite{Henheik_Teufel_2022} or only a gap in infinite volume but uniqueness of the infinite volume ground state \cite{henheik2022adiabatic}.

In the present work we can do without the uniqueness assumption from \cite{henheik2022adiabatic} and at the same time also weaken the decay assumptions for all involved operators from exponential decay to super-polynomial decay.
The fact that uniqueness does not have to be assumed is an important step, because even in simple models such as the Ising model, uniqueness can fail to hold. Furthermore, even if the ground state is expected to be unique, proving the uniqueness of an infinite-volume ground state is a formidable task. As far as we know, for fermionic systems uniqueness is known to hold only for non-interacting  systems.
Although weakening the decay assumption may seem like a minor improvement, the space $B_\infty$ of super-polynomially decaying interactions is the natural and presumably also optimal space in which such an adiabatic theorem can hold to all orders. It is natural because, even when assuming that $H_t$ and $H^1_t$   have finite range, the generator $S^{\varepsilon,\eta}_t$ of the dressing transformation still ends up in $B_\infty$. Last but not least,  the proof of our result does not use finite volume approximations and is therefore, as we believe, more transparent than previous proofs of the  many-body adiabatic theorem in infinite volume. 

The result and proof we formulate is for lattice fermion systems, but it also applies directly to quantum spin systems. However, it should be noted that fermions pose additional, quite subtle difficulties compared to quantum spin systems and our proof relies on two recent results for lattice fermion systems: The uniqueness assumption  was required in \cite{henheik2022adiabatic} because the proof of automorphic equivalence of \cite{moon2019automorphic}, an essential ingredient to proving the adiabatic theorem, generalises to lattice fermions only under the additional assumption that the ground state is gauge-invariant, which in \cite{henheik2022adiabatic} was concluded from uniqueness. In~\cite{Becker_Teufel_Wesle_2025}, we recently proved the automorphic equivalence of gapped ground states for lattice fermions, without making the assumption of gauge-invariance. See Remark~1.2 in \cite{Becker_Teufel_Wesle_2025} for details.
In order to transition from exponentially decaying interactions to $B_\infty$-interactions, we require   appropriate Lieb–Robinson bounds with algebraic light cones. Such bounds are known for both  quantum spin systems \cite{MKN2016}, \cite{EMNY2020} and lattice fermions \cite{TeufelWessel25}. As part of our assumptions, we formulate a parameter-dependent Lieb–Robinson bound and remark that recent improved Lieb–Robinson bounds for fermions with long-range interactions \cite{TeufelWessel25} imply that this assumption is satisfied for a certain parameter range. Should better Lieb–Robinson bounds become available in future, this would also lead to improved bounds in our adiabatic theorem.

The rest of the paper is structured as follows: In Section~\ref{sec:setup} we describe the mathematical setup. In Section~\ref{sec:main} we formulate and discuss the generalized adiabatic theorem, our main result. And its proof is given in Section~\ref{Section: Proof}, where several technical lemmas have been moved to appendices.

\section{Mathematical Setup} \label{sec:setup}

In this section, we will explain the basic mathematical framework, within which the main theorem is proven. We will freely refer to the objects defined here throughout the rest of the document. Most of it is a collection of standard notions used to describe interacting lattice fermions with some non-standard additions.

\subsection{The CAR-algebra and subspaces of quasi-local operators}

The anti-symmetric  (or fermionic) Fock space over the lattice $\Z^d$ is  
\begin{align*}
   \mathcal{F}(\Z^d,\C^n) \coloneq \bigoplus_{N=0}^{\infty}\ell^2(\Z^d,\C^n)^{\wedge N}.
\end{align*}
We use $a^*_{x,i}$ and $a_{x,i}$ for $x\in \Z^d$, $i\in \{1,\dots,n\}$, to denote the fermionic creation and annihilation operators associated to the standard basis of $\ell^2(\Z^d,\C^n)$ and recall that they satisfy the canonical anti-commutation relations (CAR).
The number operator at site $x\in\Z^d$ is defined by
\begin{align*}
    n_x \coloneq \sum_{i=1}^n a^*_{x,i}a_{x,i}\,.
\end{align*}
The algebra of all bounded operators on $\mathcal{F}(\Z^d,\C^n)$ is denoted by $\mathcal{B}(\mathcal{F}(\Z^d,\C^n))$. 
For each $M\subseteq \Z^d$ let $\mA_M$ be the C$^*$-subalgebra of $\mathcal{B}(\mathcal{F}(\Z^d,\C^n))$ generated by
\begin{align*}
    \{a^*_{x,i}~|~x\in M,~ i\in \{1,\dots,n\}\}~.
\end{align*}
The C*-algebra $\mA \coloneq \mA_{\Z^d}$ is the CAR-algebra, which we also call the quasi-local algebra. We write $P_0(\Z^d) \coloneq \{M\subseteq \Z^d~|~|M|<\infty \}$ and call
\begin{align*}
    \mA_0 \coloneq \bigcup_{M\in P_0(\Z^d)} \mA_M \subseteq \mA
\end{align*}
the local algebra. Consequently, an operator is called quasi-local if it lies in $\mA$ and local if it lies in $\mA_0$.
For each $\varphi \in \R$ there is a unique automorphism\footnote{In the following the term {\it automorphism} 
is used in the sense of a $*$-automorphism as defined for example in~\cite{bratteliI}.}
$g_\varphi$ of $\mA$, such that
\begin{align*}
    g_\varphi(a^*_{x,i}) =  \e^{i\varphi} \, a^*_{x,i}, \quad \text{for all}~~ x\in \Z^d,~ i\in \{ 1,\dots,n \} ~.
\end{align*}
One defines the set of even quasi-local operators
\begin{align*}
    \mA^+ \coloneq \{A\in \mA~|~   g_\pi (A) = A \}
\end{align*}
and the set of gauge-invariant quasi-local operators
\begin{align*}
    \mA^N \coloneq \{A\in \mA~|~  \forall \varphi \in \R :\, g_\varphi (A) = A \} \, .
\end{align*}
Their parts in $M\subseteq\Z^d$  are denoted by $\mA^+_M \coloneq  \mA^+\cap \mA_M$ and $\mA^N_M \coloneq \mA^N\cap \mA_M$. Note that gauge-invariant operators are always even.
For disjoint regions $M_1,M_2 \subseteq\Z^d$,  $M_1\cap M_2 = \emptyset$, operators  $A\in \mA_{M_1}^+$ and $B\in \mA_{M_2}$ commute,    $[A,B] = 0$. 

Positive linear functionals of the quasi-local algebra $\w \colon \mA \to \C$ of norm $1$ are called states. In order to define quantitative notions of localization for quasi-local operators, one makes use of the fact that one can localize operators to given regions by means of the fermionic conditional expectation. 
To this end first note that  $\mA$ has a unique state $\w^{\tr}$ that satisfies
\begin{align*}
    \w^{\tr}(AB) = \w^{\tr}(BA)
\end{align*}
for all  $A,B \in \mA$, called the tracial state (e.g.\ \cite[Definition 4.1, Remark 2]{ArakiMoriya2003}).

\medskip

\begin{proposition}[{\cite[Theorem 4.7]{ArakiMoriya2003},\cite[Proposition 2.1]{wmmmt2024}}]\label{Ex+UniqueExpectation}
    For each $M\subseteq\Z^d$ there exists a unique linear map 
    \begin{align*}
        \E_M:\mA \to \mA_M\,,
    \end{align*}
    called the conditional expectation with respect to $\w^\tr$, such that
    \begin{equation}\label{eq:Conditional expectation defining property}
      \forall A\in \mA \; \;\forall B\in \mA_M \,:\quad \w^{\tr}(AB)=\w^{\tr}(\E_M(A)B) \,.
    \end{equation}
    It is unital, positive and has the properties 
    \begin{eqnarray*}
        \forall M\subseteq \Z^d\;\; \forall A,C\in \mA_M\;\; \forall B\in\mA\, : &&   \E_M \br{A\,B\,C} = A\, \E_M(B)\,C\\[1mm]
        \forall M_1,M_2 \subseteq \Z^d\,:&&   \E_{M_1} \circ \E_{M_2}  = \E_{M_1\cap M_2}\,\\[1mm]
        \forall M \subseteq \Z^d\,:&& \E_M \mA^+ \subseteq \mA^+  \quad\mbox{and}\quad \E_M \mA^N \subseteq \mA^N \\[1mm]
        \forall M \subseteq \Z^d \; \;  \forall A \subseteq \mA\,:&&  \norm{\E_M(A)}\leq \norm{A}\,.
    \end{eqnarray*}
\end{proposition}
With the help of $\E$ we can define subspaces of $\mA$ that contain operators with well-defined decay properties.

\medskip

\begin{definition}\label{def:norm}
    For $\nu \in \N_0$, $x \in \Z^d$ and $A \in \mA$ let
    \begin{align*}
        \norm{A}_{\nu,x} \coloneq \norm{A} + \sup_{k\in \N_0} \norm{A-\E_{B_k(x)}A} \, (1+k)^\nu \, ,
    \end{align*}
    where $B_k(x) \coloneq \{y\in \Z^d \, | \, \norm{x-y} \leq k \}$ is the box with side-length $2k$ around $x$ with respect to the maximum norm on $\Z^d$.
    We denote the set of all $A\in \mA$ with   $\norm{A}_{\nu,x}<\infty$ for some (and therefore all) $x\in \Z^d$ and all $\nu\in \N_0$ by $\mA_\infty$. 
    Somewhat abusing notation, we abbreviate for $A\in\mA_\infty$
    \[
    \norm{A}_\nu := \min_{x\in\Z^d}\norm{A}_{\nu,x}\,.
    \]
    Note that $\|\cdot\|_\nu:\mA_\infty \to [0,\infty)$ is not a norm, as it does not satisfy the triangle inequality: the sum of two operators that are well localized at far away points is no longer well localized. Still, it turns out to be a useful tool for quantifying   the localization of operators in a translation invariant way.
\end{definition}

The following lemma justifies that the conditional expectation of a quasi-local operator approximates it locally.

\medskip
\begin{lemma} \label{lem: limit in decay norm}
    Let $\nu \in \N_0$, $x \in \Z^d$ and $A \in \mA_\infty$. Then
    \begin{align*}
        \lim_{k \to \infty} \lVert A - \E_{B_k(x)} \, A \rVert_{\nu, x} = 0 \, .
    \end{align*}
\end{lemma}
\begin{proof}
    Let $\Z^d \to \aut(\mA)$, $\gamma \mapsto T_\gamma$ be the standard translation on $\mA$. Then 
    \begin{align*}
        \lVert A - \E_{B_k(x)} \, A \rVert_{\nu , x} &= \lVert T_{-x} \, A - T_{-x} \, \E_{B_k(x)} \, A \rVert_{\nu , 0}
        \\
        & = \lVert T_{-x} \, A - \E_{B_k(0)} \, T_{-x} \, A \rVert_{\nu , 0} \, .
    \end{align*}
    The statement now follows with \cite[Lemma 2.3]{wmmmt2024}, since the additionally required gauge-invariance was not used in its proof.
\end{proof}

The next lemma quantifies the idea that, even if their supports are not strictly disjoint, operators localized in different regions have small commutators.

\medskip

\begin{lemma}[{\cite[Lemma 2.3]{Becker_Teufel_Wesle_2025}}] \label{lem:commutator bound}
    Let $A,B \in \mA_\infty$, such that either $A$ or $B$ is even, $\nu,m \in \N_0$ and $x,y \in \Z^d$. It holds that
    \begin{align*}
        \norm{[A,B]}_{\nu,x} \leq 4^{\nu+m+3} \frac{\norm{A}_{\nu+m,y}\, \norm{B}_{\nu+m,x}}{(1+\norm{x-y})^m}\, .
    \end{align*}
    Here the norm on $\Z^d$ is again the maximum norm.
\end{lemma}

\subsection{Spaces of interactions and induced derivations}

The relevant physical dynamics on $\mA$ is generated by densely defined derivations, which in turn are constructed from so-called interactions. More generally interactions are used to model extensive quantities on the infinite lattice.
An interaction is a map $\Phi: P_0(\Z^d) \to \mA^N$, such that $\Phi(\emptyset) = 0$ and for all $M\in P_0(\Z^d)$ it holds that $\Phi(M) \in \mA_M$, $\Phi(M)^* = \Phi(M)$, and the sum 
\begin{equation}\label{eq:uncond}
    \sum_{\substack{K\in P_0(\Z^d)\\ M\cap K \neq \emptyset}} \Phi(K)
\end{equation}
converges unconditionally.  Note that in our definition the local terms of an interaction are always gauge-invariant. 
Interactions  define derivations on $\mA$ in the following way:    
For an interaction $\Phi$ let
\begin{align*}
    \mL_{\Phi}^\circ: \mA_0 \to \mA,~ A\mapsto \sum_{M\in P_0(\Z^d)} [\Phi(M),A] \, .
\end{align*}
It follows from \cite[Proposition 3.2.22]{bratteliI} and \cite[Proposition 3.1.15]{bratteliI} that $\mL_{\Phi}^\circ$ is closable. We denote its closure by $\mL_{\Phi}$ and call it the Liouvillian of $\Phi$.

Below we define several spaces of well-behaved  interactions that will be used later.

\medskip

\begin{definition}\label{norm interaction}
    For an interaction $\Phi$   and $\nu \in \N_0$ let
    \begin{align*}
        \|\Phi\|_{\nu} \coloneq \sup_{x\in \Z^d} \sum_{\substack{M\in
        P_0(\Z^d)\\x\in M}}(1+\mathrm{diam}(M))^\nu  \| \Phi(M)\|  ~.
    \end{align*}
    The set of interactions with finite $\norm{\cdot}_{\nu}$ for all $\nu \in \N_0$ is denoted by $B_{\infty}$. 
    By $\mathrm{diam}(M)$ we mean the maximal distance of two elements in $M$ with respect to the maximum norm on $\Z^d$.\\
    Let $I\subseteq \R$ be an interval. Define $B_{\infty,I}^{(k)}$ as the set of families of $B_\infty$-interactions $(\Phi_s)_{s \in I}$, where for each $M$ the map $s \mapsto \Phi_s(M)$ is  $k$ times continuously differentiable and that satisfy $\sup_{s \in I}\norm{\frac{\dd^m}{\dd s^m}\Phi_s}_{\nu} < \infty$ for all $\nu \in \N_0$ and $m\leq k$.
\end{definition}

\medskip

\begin{definition}\label{def:lipschitz potential}
    A Lipschitz potential $V$ is an on-site interaction, i.e.\ only supported on one-element sets, such that for all $x\in \Z^d$ it holds that $V(\{x\}) = v(x)\, n_x$, where $v:\Z^d \to \R$ is a function that satisfies 
    \begin{align*}
     \exists\, C_v\in\R \quad\ \forall x,y\in\Z^d\,:\quad    \lvert v(x)- v(y) \rvert  \leq C_v\,\lVert x- y \rVert\,.
    \end{align*}
     Any $C_v$ that satisfies the condition is called a Lipschitz constant for $V$. We denote the set of all Lipschitz potentials by $\mV$.\\
    Let $I \subseteq \R$ be an interval. Define $\mV_I^{(k)}$ as the set of families of Lipschitz interactions $(V_s)_{s \in I}$ with uniform Lipschitz constant that are term-wise $k$ times continuously differentiable and for each $m\leq k$ it holds that $(\frac{\d^m}{\d s^m} \, V_s)_{s\in I}$ is again a family of Lipschitz potentials with uniform Lipschitz constant.
\end{definition}

\medskip

\begin{definition}
Let $L\subseteq \Z^d$. We say that an interaction $\Phi$ is $L$-localized, if for all $n,\nu\in\N_0$
\[
\sup_{x\in\Z^d}\sup_{A\in\mA_\infty^N\setminus\{0\}} \frac{
\|\mL_\Phi A\|_{\nu,x}\,(1 + \d(x,L))^n} {\|A\|_{\nu+n+d+2,x}}<\infty\,.
\]
We write $B_{\infty,L}$ for $L$-localized $B_\infty$-interactions and $\mV_L$ for $L$-localized Lipschitz potentials.
\end{definition}

\medskip

\begin{remark}
 Note that Lemma~\ref{lem:commutator bound} and Lemma~\ref{lem: sum representation of generator} imply that a sufficient condition for $\Phi\in B_\infty$ being in $B_{\infty,L}$ is super-polynomial decay of the local terms $\Phi_x$ away from $L$, i.e.\  
 \[
 \sup_{x\in\Z^d} \|\Phi_x\|_{\nu,x} (1 + \d(x,L))^n <\infty
 \]
 for all $n,\nu\in\N_0$. A sufficient condition for $V\in\mV$ to be in $\mV_L$ is that it is constant on every connected component of $\Z^d\setminus L$ (understood as a subset of the graph $\Z^d$).
\end{remark}

\medskip

\begin{definition}
    Let $(\Phi_s)_{s \in I} \in B_{\infty, I}^{(k)}$ be such that $\frac{\d^m }{ \d s^m} \, \Phi_s \in B_{\infty, L}$ for all $s \in I$ and $m \leq k$ such that the $L$-localization is uniform in $s \in I$, then we write $(\Phi_s)_{s \in I} \in B_{\infty, I, L}^{(k)}$.\\
    Likewise, if $(V_s)_{s \in I} \in \mV_I^{(k)}$ is such that $\frac{\d^m}{\d s^m} \, V_s \in \mV_L$ for all $s \in I$ and $m \leq k$, uniformly in $s \in I$, then we write $(V_s)_{s \in I} \in \mV_{I, L}^{(k)}$.
\end{definition}

\medskip

For two interactions $\Phi$ and $\Psi$, we define their commutator to be the map
\begin{equation}
    [\Phi, \, \Psi] : P_0(\Z^d) \to \mA^N \; , \; M \mapsto [\Phi, \, \Psi](M) \coloneq \sum_{\substack{M_1, M_2 \subseteq M \\ M_1 \cup M_2 = M}} [\Phi(M_1), \, \Psi(M_2)] \, .
\end{equation}
It is easy to check that $\i [\Phi,\Psi]$ satisfies all the conditions of the above definition of interactions except unconditional convergence. All commutators in the following will however be interactions:

\medskip

\begin{restatable}[label=lem: comm is smooth]{lemma}{smoothcommutator}
      Let $k \geq 0$ and $L \subseteq \Z^d$ be some region. Assume that $(\Phi_t)_{t \in I}  \in B_{\infty, I, L}^{(k)}$, $(V_t)_{t \in I} \in \mV_{I, L}^{(k)}$, and $(\Psi_t)_{t \in I} \in B_{\infty, I}^{(k)}$. Then it holds that
    \begin{align*}
        (\i [ \Phi_t + V_t, \, \Psi_t])_{t \in I} \in B_{\infty, I, L}^{(k)} \, .
    \end{align*}
\end{restatable}
\hyperref[proof:smoothcommutator]{%
    The proof of Lemma~\ref*{lem: comm is smooth}%
} is given in Appendix~\ref{sec: techincal lemmas}.

\medskip

One can sum all the local terms of an interaction that are associated to one lattice point to obtain a quasi-local observable. In computations, it is convenient to work with this family of quasi-local terms, also called a $0$-chain.

\medskip

\begin{definition}\label{def: quasi-local terms}
    For $M \in P_0(\Z^d)$ we define its center $\mathrm{C}(M) \in M$ as the point in $M$ that minimizes the distance to its center of mass $\mathrm{cm}(M)\in\R^d$. If there are multiple such points we choose the one where the standard polar-coordinate angles $$\angle(\mathrm{C}(M)-\mathrm{cm}(M))\in [0,\pi)^{d-2}\times [0,2\pi)$$ are minimal with respect to lexicographical ordering, if $d\geq 2 $ and the one larger than $\mathrm{cm}(M)$ if $d=1$.
    Let $x\in \Z^d$. We define $R_x \subseteq P_0(\Z^d)$ to be set of all finite subsets of $\Z^d$ that have their center in $x$. Given an interaction $\Phi$ we define 
    \begin{align*}
        \Phi_x \coloneq \sum_{M \in R_x} \Phi(M) \, .
    \end{align*}
\end{definition}

\medskip

\begin{remark} \label{rem: bound zero chain}
    Note that for any interaction $\Phi$ and lattice point $x$ the quasi local observable $\Phi_x$ is well-defined and that $ \lVert \Phi_x \rVert_{\nu,x} \leq 3 \, \lVert \Phi \rVert_\nu $ for all $\nu \in \N_0$.
\end{remark}

\medskip

\begin{restatable}[label=lem: sum representation of generator]{lemma}{sumrepresentation}
      Let $\Phi \in  B_\infty$ and $V \in \mV$ be a Lipschitz potential associated to the Lipschitz function $v\colon \Z^d \to \R$ with Lipschitz constant $C_v$ . It holds that $\mA_\infty \subseteq D(\mL_{\Phi +V})$. For all $ A\in \mA_\infty$ the sums
    \[\sum_{M\in P_0(\Z^d)} [\,\Phi(M) + V(M)  , \,A\,] \quad \text{and} \quad \sum_{x\in \Z^d} [\, \Phi_x + V_x, \, A \, ] \]
    converge absolutely and
    \[ \mL_{\Phi + V } \, A 
    \;=\;  \sum_{M\in P_0(\Z^d)} [\,\Phi(M) + V(M)  , \,A\,]
    \;=\;  \sum_{x\in \Z^d} [\, \Phi_x + V_x, \, A \, ] 
    \, .
    \]
    For each $\nu \in \N_0$ there is a constant $c_\nu$, independent of $\Phi$, $V$ and $ A $, such that for all~$x\in \Z^d$
    \[
    \norm{ \mL_{\Phi + V } \, A }_{\nu,x} \leq c_\nu \, \big( \norm{\Phi}_{\nu+d+1} \, \norm{A}_{\nu+d+1,x} +  (C_v + \lvert v(x) \rvert )\,   \norm{A}_{\nu+d+2,x} \big)
    \,.
    \]
    If $A$ is gauge-invariant then, with the same constant as above one gets
    \begin{align*}
        \norm{ \mL_{\Phi + V } \, A }_{\nu,x} \leq c_\nu \, \big(  \norm{\Phi}_{\nu+d+1} \, \norm{A}_{\nu+d+1,x} +  C_v\,   \norm{A}_{\nu+d+2,x} \big)
    \end{align*}
\end{restatable}
\hyperref[proof:sumrepresentation]{%
    The proof of Lemma~\ref*{lem: sum representation of generator}%
}  is given in Appendix~\ref{sec: techincal lemmas}.

\medskip

In the infinite-volume setting the usual spectral gap condition is replaced by the following algebraic one. It implies in particular that the associated GNS-Hamiltonian has a unique gapped ground state in the usual sense, see \cite[Theorem A.3]{tasaki2022lieb}.

\medskip

\begin{definition}\label{def:gap}
    A state $\w$ is called a \emph{ground state} of an interaction $\Phi$, if for all $A \in D(\mL_\Phi)$
    \begin{align*}
        \w(A^*\, \mL_\Phi \, A) \geq 0 \, .       
    \end{align*}
    It is called a \emph{gapped ground state} of $\Phi$ with gap $g>0$, if for all $A \in D(\mL_\Phi)$
    \begin{align*}
        \w(A^*\, \mL_\Phi \, A) \geq g \, (\w( A^*A) - \lvert \w(A) \rvert^2)\, .
    \end{align*}
\end{definition}

\subsection{Cocycles of automorphisms and Lieb--Robinson bounds}

Well-behaved families of interactions generate cocylces of automorphisms on the quasi-local algebra, which describe evolutions in the infinite-volume setting.

\medskip

\begin{definition}
    Let $I\subseteq \R$ be an interval. A family $(\alpha_{u,v})_{(u,v) \in I^2}$ of automorphisms of $\mA$ is called a \emph{cocycle} if it satisfies
    \begin{align*}
        \forall t,u,v \in I, \quad \alpha_{t,u} \, \alpha_{u,v} =\alpha_{t,v}\, .
    \end{align*}
    We say the cocycle is generated by the family of interactions $(\Phi_{v})_{v\in I}$, if for all $A \in \mA_0$ it holds that 
    \begin{align*}
        \partial_v \, \alpha_{u,v} \, A = \alpha_{u,v} \, \i \, \mL_{\Phi_{v}} \, A  \, .   
    \end{align*}
    We call $(\alpha_{u,v})_{(u,v) \in I^2}$ locally generated if it is generated by some family of interactions in $B_{\infty,I}^{(0)}$.
\end{definition}

\medskip

\begin{lemma}\label{lem: cocycles}
    Let $I \subseteq \R$ be an interval and $(\Phi_{v})_{v\in I}$ a family of interactions in $B_{\infty,I}^{(0)}$ and $(V_v)_{v \in I}$ a family of Lipschitz potentials in $\mV_I^{(0)}$. Then there exists a unique cocycle of automorphisms $(\alpha_{u,v})_{(u,v) \in I^2}$ generated by $(\Phi_{v} + V_v)_{v\in I}$ and the following Lieb--Robinson bound holds: 
    
    For each  $\nu \in \N_0$ there exist constants $C_\nu, c_\nu>0$, such that for all finite disjoint sets $X,Y \subseteq \Z^d$ and all $A \in \mA_X^+$, $B\in \mA_Y$, it holds that
    \begin{align*}
        \lVert [\, \alpha_{u,v} \, A, \, B \,] \rVert 
        \leq  \lVert A \rVert \, \lVert B \rVert \, \lvert Y \rvert \,  \frac{ C_\nu\, (1+ |v-u|^{2d+1})) }{ ( 1 + \max( 0, \mathrm{dist}(X,Y)^{1/2} - c_\nu \, \lvert v - u \rvert) )^\nu }  \, .
    \end{align*}
    The dependence on $(\Phi_v)_{v\in I}$ of $C_\nu$ and $c_\nu$ is such that for each $C>0$ they can be chosen uniformly for all $(\Phi_v)_{v\in I}$ with $\sup_{v\in I} \lVert \Phi_v \rVert_{2\nu+d} < C$.
    
    Furthermore, the above bound implies that for all $\nu \in \N_0$ there exists an increasing function $b_\nu:\R \to \R$, that grows at most polynomially, such that
      for all  $x\in \Z^d$, $A\in \mA_\infty$, and  $ u,v\in I$
    \begin{align*}
        \norm{\alpha_{u,v} \, A}_{\nu,x} \leq b_\nu(|v-u|) \norm{A}_{\nu,x} \,,
    \end{align*}
    where for all  constants $C>0$ the function $b_\nu$ can be chosen uniformly for all $(\Phi_v)_{v\in I}$ \linebreak with $\sup_{s\in I} \lVert \Phi_v \rVert_{4\nu+9d+4} < C$.  
\end{lemma}
\begin{proof}
    The case for $B_\infty$-interactions without a Lipschitz potential is covered in \cite[Lemma 2.10]{Becker_Teufel_Wesle_2025} and \cite[Proposition A.1]{Becker_Teufel_Wesle_2025}. The existence and uniqueness of the automorphism is ensured by \cite[Corollary 5.2]{Bru2017}, which also applies to Lipschitz potentials. The Lieb--Robinson bound of \cite[Theorem 6]{TeufelWessel25}, which is used to prove the corresponding statements in \cite{Becker_Teufel_Wesle_2025}, is insensitive to on-site Hamiltonians. Hence, the statement also holds for cocycles of automorphisms generated by $B_\infty$-interactions with added Lipschitz potentials. 
\end{proof}

\section{The Generalized Adiabatic Theorem}\label{sec:main}

Let $I \subseteq \R$ be some interval and $L \subseteq \Z^d$ some (not necessarily bounded) region. We are concerned with the dynamics generated  by the Hamiltonian  $
    H^{\varepsilon, \eta}_t \coloneq \frac{1}{\eta} \, (H_t + \eps \, (H^1_t + V_t))$.
    Here $(H_t)_{t \in I}$ is a time-dependent  Hamiltonian with gapped ground state $(\w_t)_{t \in I}$, $H_t^1$ is a time-dependent bounded interaction, and $V_t$ a time-dependent Lipschitz potential.
The adiabatic parameter  $\eta \in (0,1]$ controls the rate at which the operators vary in time and $\eta\ll 1$ is called the adiabatic regime of slow variation. 
We call $\eps   (H^1_t + V_t)$ the perturbation and $\eps \in [0,1]$ controls its strength. 
Our main theorem holds under the following assumptions on these objects:

\paragraph{Assumptions on the Interactions} 
\begin{enumerate}[label=(I\arabic*),ref=I\arabic*]
    \item \label{ass: diff. of interactions}
    \textbf{Regularity of interactions:}
    The family of interactions $(H_t)_{t \in I}$ is in $B_{\infty, I}^{(\infty)}$ and the perturbations are localized in $L$ in the sense that $(\Dot{H}_t)_{t \in I}$ and $(H^1_t)_{t \in I}$ lie in $B_{\infty, I, L}^{(\infty)}$ and $(V_t)_{t \in I}$ lies in $\mV_{I, L}^{(\infty)}$. 
\end{enumerate}
According to Lemma~\ref{lem: cocycles},   Assumption~\eqref{ass: diff. of interactions} implies that  for $\eta\in(0,1]$ and $\eps\in[0,1]$ the family of interactions 
\[H^{\varepsilon, \eta}_t \coloneq \tfrac{1}{\eta} \, (H_t + \eps \, (H^1_t + V_t))
    \] generates a cocycle of automorphisms of $\mA$ which we denote by 
 $(\mU_{s, t}^{\varepsilon, \eta})_{(s,t)\in I^2}$.

\begin{enumerate}[label=(I\arabic*),ref=I\arabic*]
\addtocounter{enumi}{1}
    \item \label{ass: LR bound} 
     \textbf{Lieb-Robinson bound:}
    There exist constants $\kappa \geq 1$,  $\nu \geq \kappa \, (d + 1)$, and $v\geq 0$, and an at most polynomially growing function $f : \R_+ \to \R_+$ such that for all disjoint sets $X, \, Y \subseteq \Z^d$, all $A \in \mA_X^+, \, B \in \mA_Y$, and all $s, t \in I$, $\eps \in [0,1]$, $\eta \in (0,1]$ it holds that 
    \begin{align*}
        \lVert [\mU_{s, t}^{\varepsilon, \eta} \, A, \, B] \rVert \leq \lVert A \rVert \, \lVert B \rVert \, \lvert Y \rvert \, \frac{f( \tfrac{1}{\eta}\lvert t - s \rvert)}{\left(1+ \max (0, \mathrm{dist}(X, \, Y)^{1/\kappa} -  \tfrac{v }{\eta} \, \lvert t - s \rvert)\right)^\nu} \, .
    \end{align*} 
\end{enumerate}

According to Lemma~\ref{lem: cocycles},   Assumption~\eqref{ass: diff. of interactions}
    implies \eqref{ass: LR bound} for   $\kappa=2$ and $f(t)= C(1+t^{2d+1})$. Actually, the proof in \cite{Becker_Teufel_Wesle_2025} generalises to any $\kappa>1$. We chose to add \eqref{ass: LR bound} as an explicit assumption nonetheless, because we expect that  \eqref{ass: diff. of interactions} implies \eqref{ass: LR bound} with even better exponents. For example,   in case the interactions $H_t$ and $H_t^1$ decay exponentially, standard Lieb-Robinson bounds imply \eqref{ass: LR bound} with $\kappa=1$ and $f(t)\equiv C$.

\paragraph{Assumptions on the Ground States}
\begin{enumerate}[label=(G\arabic*),ref=G\arabic*]
    \item \label{ass: gap of gs} 
     \textbf{Gapped ground state:} There exists a constant $g>0$, such that for each $t\in I$, the state  $\w_t$ is a   gapped ground state of $H_t$ with gap at least $g$ (cf.\ Definition~\ref{def:gap}).
    \item \label{ass: diff. of gs} 
     \textbf{Regularity of the ground state:}
     For all $A \in \mA_\infty$ the map $t \mapsto \w_t(A)$ is differentiable  and there is a $\nu \in \N_0$, such that for some (and therefore all) $x \in  \Z^d$, there exists a constant $C_{\nu,x}$ such that for all $ A \in \mA_\infty$ and $t\in I$
    \begin{align*}
        |\Dot{\w}_t(A)|\leq C_{\nu,x} \, \lVert A\rVert_{\nu,x} \, .
    \end{align*}
\end{enumerate}

Let us briefly comment on the differences of our assumptions to those of the closest existing result from 
\cite{henheik2022adiabatic}. First, in contrast to \cite{henheik2022adiabatic},  we do not assume that $\w_t$ is the unique ground state of $H_t$ and also our assumptions on the differentiability of the map $t \mapsto H_t$ are weaker. Secondly, we assume that the families of interactions $H_t$ and $H_t^1$ have super-polynomial rather than exponential decay.  And last but not least, we keep track of the region $L\subseteq\Z^d$ in which the Hamiltonian changes and in which the perturbations are localized. This feature is used, for example, in the analysis of Hall currents along potential steps in \cite{TW2025}.

\medskip
\begin{theorem}\label{Theorem: main theorem} {\bf (Generalized adiabatic theorem)}\; 
Let $I \subseteq \R$ be an interval, $L\subseteq \Z^d$ a region, and let $(H_t)_{t \in I}$, $(H^1_t)_{t \in I}$, $(V_t)_{t \in I}$ and $(\w_t)_{t \in I}$ satisfy \eqref{ass: diff. of interactions}, \eqref{ass: LR bound}, \eqref{ass: gap of gs}, and \eqref{ass: diff. of gs}. \\[1mm]
There exists a two-parameter family of time-dependent interactions
\[
[0,1]^2\ni(\eps,\eta)\mapsto (S_t^{\varepsilon, \eta})_{t \in I} \in B_{\infty, I, L}^{(\infty)}\,,
\]
which  is uniformly bounded in the sense that $\sup_{(\eps,\eta) \in [0, 1]^2} \,  \sup_{t \in I} \, \lVert S_t^{\varepsilon, \eta} \rVert_\nu < \infty$ for all $\nu \in \N_0$, and
 generates   automorphisms $\beta_t^{\varepsilon, \eta} := \exp ( \i  \mL_{S_t^{\varepsilon, \eta}})$,    such that the super-adiabatic states   
\begin{align*}
    \w_t^{\varepsilon, \eta} \coloneq \w_t \circ \beta_t^{\varepsilon, \eta}  
\end{align*}
almost intertwine the time evolution $\mU_{t_0, t}^{\varepsilon, \eta} $ in the following sense:\\[1mm] 
For every $n \in \N$ there exists a constant $C_n$ such that for all $ t, \, t_0 \in I$, $\eps \in [0, \, 1]$, $\eta  \in (0, \, 1]$, $M \in P_0(\Z^d)$, and $A \in \mA_M$  
\begin{align*}
    \left|\w_t^{\varepsilon, \eta}(A) - \w_{t_0}^{\varepsilon, \eta} (\mU_{t_0, t}^{\varepsilon, \eta} \, A) \right| 
    \;\leq\; C_n   \left(\varepsilon^{n + 1} + \eta^{n + 1} \right) \tfrac{|t - t_0|}{\eta} \, \left( 1 + \left(\tfrac{|t - t_0|}{\eta}\right)^{\kappa d} \, f \left( \tfrac{\lvert t - t_0 \rvert}{\eta}  \right) \right) \, \lVert A \rVert \, |M|^2 \, .
\end{align*}
The statement holds also for $A \in \mA_\infty$, if we replace $\lVert A \rVert \, \lvert M \rvert^2$ by $ \lVert A \rVert_{2d+2}$ (with a different constant  $C_n$).

In addition, the following holds:
\begin{enumerate}
        \item \textup{\textbf{Locality of $S_t^{\epsi,\eta}$ in time and space:}} For each $t \in I$ the generator $S_t^{\epsi,\eta}$ of the super-adiabatic automorphism $\beta_t^{\varepsilon, \eta}$ is determined by $H_t$, $H^1_t$, $V_t $ and their time derivatives at time $t$. Its dependence on $\dot H_t$ and $H^1_t$ and $V_t$ is  quasi-local in space in the sense that for all $m, \nu \in \N_0$ is holds that
        \begin{align*}
            \sup_{(\eps,\eta) \in [0, 1]^2}  \, \sup_{t \in I} \,  \sup_{x \in \Z^d} \, \lVert (S_t^{\varepsilon, \eta})_x \rVert_{\nu, x} \, (1 + \d (x, L))^m < \infty \, .
        \end{align*}
        \item \textup{\textbf{Stationarity:}} If on some interval $J\subseteq I$ the Hamiltonian $H_t^{\varepsilon,\eta}\equiv H^{\epsi,\eta}_J$ is constant,  then  the generator 
        $S_t^{\epsi,\eta}\equiv S_J^{\epsi,0}$ and thus   the super-adiabatic state 
        $\w_t^{\eps, \eta} \equiv \w_J^{\eps, 0}$ are constant on $J$ as well and independent of $\eta$. Moreover,  
        for all $t,t_0\in J$, $\eps\in[0,1]$,   $A\in\mA_\infty$ 
        \[
         \left|\w_J^{\varepsilon, 0}(A) - \w_J^{\varepsilon, 0} (\mU_{t_0, t}^{\varepsilon, 1} \, A) \right| \;\leq\; C_n \,\epsi^{n+1}|t - t_0|\left( 1 + |t - t_0|^{\kappa d} \, f \left(   \lvert t - t_0 \rvert \right) \right) \, \lVert A \rVert_{2d+2}\,,
        \]
        and thus, 
        \begin{align*}
             \sup_{A \in \mA_{\infty}} \frac{\lvert \w_J^{\eps, 0}(\mL_{H_J^{\eps, 1}}\, A) \rvert}{\lVert A \rVert_{2d+2}}
            = \mathcal{O}(\eps^{\infty})\,.
        \end{align*}
        In this situation we call $\w_J^{\eps,0}$ a NEASS (non-equilibrium almost-stationary state) for $ H_J^{\eps, 1} $.

        \item \textup{\textbf{Vanishing Perturbation:}} If for some $t \in I$ all time derivatives of $H_t^{\eps, \eta}$ vanish and $H_t^1=V_t=0$, then     $\w_t^{\eps, \eta} = \w_t$ for all $\eta, \, \eps \in [0,1]$.
    \end{enumerate}
\end{theorem}
\medskip

\section{Proof of   Theorem~\ref{Theorem: main theorem}} \label{Section: Proof}

The proof mainly consists of two steps. First, for each $n\in \N$ we construct an $n$-dependent family of interactions $(\Tilde{S}_t^{\varepsilon, \eta})_{t\in I}$ such that the $n$-dependent states $\Tilde{\w}_t^{\varepsilon, \eta} \coloneq \w_t \circ \exp \big(  \i  \, \mL_{\Tilde{S}_t^{\varepsilon, \eta}} \big)$ satisfy the bound 
\begin{align*}\label{eq:n-bound}
    |\Tilde{\w}_t^{\varepsilon, \eta}(A) - \Tilde{\w}_{t_0}^{\varepsilon, \eta} (\mU_{t_0, t}^{\varepsilon, \eta} \, A) | 
    \leq C_n \left(\varepsilon^{n + 1} + \eta^{n + 1} \right) \tfrac{|t - t_0|}{\eta} \, \left( 1 + \left(\tfrac{|t - t_0|}{\eta}\right)^{\kappa d} \, f \left( \tfrac{\lvert t - t_0 \rvert}{\eta}  \right) \right) \, \lVert A \rVert \, |M|^2 \, . \numberthis
\end{align*}
To obtain such interactions, we make the ansatz 
\begin{align} \label{def: S^n}
    \Tilde{S}_t^{\varepsilon, \eta} \coloneq \sum_{j = 1}^{n} \sum_{i = 0}^j \varepsilon^i \, \eta^{j - i} \, K_t^{j, i} \, ,
\end{align}
with $(K_t^{j, i})_{t \in I} \in B_{\infty, I, L}^{(\infty)}$ independent of $n$ and $\eps$ and $\eta$. We will then iteratively choose $K_t^{j, i}$ such that \eqref{eq:n-bound} is satisfied for each $n$. In the second step of the proof we construct an $n$-independent family of interactions $(S_t^{\eps,\eta})_{t\in I}$, such that the associated states $\w_t^{\varepsilon, \eta} \coloneq \w_t \circ \exp \big(  \i  \, \mL_{S_t^{\varepsilon, \eta}} \big)$ satisfy the bound claimed in the theorem. To achieve this, we use the interactions $K_t^{j, i}$ of the first step and perform a suitable `resummation' of the non-convergent series 
\begin{align*}
    \sum_{j = 1}^{\infty} \sum_{i = 0}^j \varepsilon^i \, \eta^{j - i} \, K_t^{j, i} \, ,
\end{align*}
essentially by truncating it depending on $\eps$ and $\eta$.

\subsection{Construction of the $n$-dependent generator}

We proceed with the first step of the proof. We make our ansatz \eqref{def: S^n} and for each $\lambda \in \R$ define
\begin{align*}
    \Tilde{\beta}_t^{\varepsilon, \eta} ( \lambda)
    \coloneq \exp \big(  \i  \, \lambda \, \mL_{\Tilde{S}_t^{\varepsilon, \eta}} \big)
    \, .
\end{align*}
Note that for each $A \in \mA_\infty$ the expression $\Tilde{\beta}_t^{\varepsilon, \eta} ( 1 ) \, A $ is differentiable in $t$ due to Duhamel's formula. Using the differentiability of the time evolution and of the ground state \hyperref[ass: diff. of gs]{(\ref*{ass: diff. of gs})}, we get
\begin{align*}
    \Tilde{\w}_t^{\varepsilon, \eta} (A) - \Tilde{\w}_{t_0}^{\varepsilon, \eta} (\mU_{t_0, t}^{\varepsilon, \eta} \, A) = \int_{t_0}^t \d s \, \frac{\d}{\d s} \, \w_s \left( \Tilde{\beta}_s^{\varepsilon, \eta} (1) \, \mU_{s, t}^{\varepsilon, \eta} \, A \right) \, .
\end{align*}
We apply the product rule and obtain the following three terms:
\begin{align*}
    &\Dot{\w}_s  \left( \Tilde{\beta}_s^{\varepsilon, \eta} (1) \, \mU_{s, t}^{\varepsilon, \eta} \, A \right) 
    + \w_s \left( \left( \frac{\d}{\d s}  \Tilde{\beta}_s^{\varepsilon, \eta}(1) \right) \, \mU_{s, t}^{\varepsilon, \eta} \, A \right) 
    + \w_s \left( \Tilde{\beta}_s^{\varepsilon, \eta} ( 1) \, \frac{\d}{\d s} \mU_{s, t}^{\varepsilon, \eta} \, A \right) \, .
\end{align*}
To further evaluate the first term, we use Lemma \ref{lem: automorphic equivalence}, which is a corollary of the automorphic equivalence shown in \cite{Becker_Teufel_Wesle_2025} and find
\begin{align*}
    \Dot{\w}_s \left( \Tilde{\beta}_s^{\varepsilon, \eta} (1) \, \mU_{s, t}^{\varepsilon, \eta} \, A \right) = - \i \, \w_s \left( \mL_{\mI_s(\Dot{H}_s)} \, \Tilde{\beta}_s^{\varepsilon, \eta}(1) \, \mU_{s, t}^{\varepsilon, \eta} \, A \right) \, ,
\end{align*}
where $\mI_s$ is the inverse Liouvillian, defined in Appendix \ref{Appendix: Inverse Liouvillian}.
For the second term, we use Duhamel's formula and get
\begin{align*}
    \Big( \frac{\d}{\d s}  \Tilde{\beta}_s^{\varepsilon, \eta}(1) \Big) \, \mU_{s, t}^{\varepsilon, \eta} \, A 
    = \int_0^1 \d \lambda 
    \, \Tilde{\beta}_s^{\varepsilon, \eta}(\lambda)
    \, \i \, \mL_{ \frac{\d}{\d s} \Tilde{S}_s^{\varepsilon, \eta}}
    \, \Tilde{\beta}_s^{\varepsilon, \eta} (-\lambda)
    \, \Tilde{\beta}_s^{\varepsilon, \eta}(1) \, \mU_{s, t}^{\varepsilon, \eta} \, A 
    \, . 
\end{align*}
By definition of the Heisenberg time evolution the derivative in the last term is given by
\begin{align*}
    \frac{\d}{\d s} \mU_{s, t}^{\varepsilon, \eta} \, A
    = -\i\,  \mL_{H_s^{\varepsilon, \eta}} \, \mU_{s, t}^{\varepsilon, \eta} \, A
    \, .
\end{align*}
We now combine these three terms and insert an identity into the last one. This will allow us to expand the automorphism $\Tilde{\beta}_s^{\eps, \eta}$ and iteratively fix the interactions $K_s^{j,i}$ that define it. We find
\begin{align*}
    \frac{\d}{\d s} \, \w_s \Big( \Tilde{\beta}_s^{\varepsilon, \eta}(1) \, \mU_{s, t}^{\varepsilon, \eta} \, A \Big) 
    &= - \,  \w_s \left( \i\, \mL_{\mI_s( \Dot{H}_s )} \, \Tilde{\beta}_s^{\varepsilon, \eta}(1) \, \mU_{s, t}^{\varepsilon,\eta} \, A \right) 
    \\
    & \quad +  \w_s \left( \int_0^1 \d \lambda \, \Tilde{\beta}_s^{\varepsilon, \eta}( \lambda) \, \i \,  \mL_{\frac{\d}{\d s} \Tilde{S}_s^{\varepsilon, \eta}} \, \Tilde{\beta}_s^{\varepsilon, \eta}( -\lambda) \, \Tilde{\beta}_s^{\varepsilon, \eta}(1) \, \mU_{s, t}^{\varepsilon, \eta} \, A \right)
    \\
    & \quad  - \left. \omega_s \left( \Tilde{\beta}_s^{\varepsilon, \eta}(\mu) \, \i 
    \, \mL_{H_s^{\varepsilon, \eta}} \, \Tilde{\beta}_s^{\varepsilon, \eta}(-\mu) \, \Tilde{\beta}_s^{\varepsilon, \eta}(1) \, \mU_{s, t}^{\varepsilon, \eta} \, A \right) \right|_{\mu = 1} \, .
\end{align*}
To simplify the notation, we set $B \coloneq \Tilde{\beta}_s^{\varepsilon, \eta} (1) \, \mU_{s, t}^{\varepsilon, \eta} \, A$, which lies in $\mA_\infty$ by Lemma \ref{lem: cocycles}. 

In the next step, we Taylor-expand the last two terms of the previous expression in the parameters $\lambda$ and $\mu$ around zero.
For the second term, we follow the proof of \cite[Proposition 3.3]{wmmmt2024} and use that $\int_0^1 \lambda^k \, \d \lambda = \frac{1}{k + 1}$. We find 
\begin{align*}
    \hspace{2em}&\hspace{-2em} \int_0^1 \d \lambda
    \, \Tilde{\beta}_s^{\varepsilon,\eta} (\lambda)
    \, \i \, \mL_{ \frac{\d}{\d s}\Tilde{S}_s^{\varepsilon, \eta}}
    \, \Tilde{\beta}_s^{\varepsilon,\eta} (-\lambda) \, B 
    \\
    & = \sum_{k = 0}^n \int_0^1 \d \lambda \, \frac{\lambda^k}{k!} \, \frac{\d^k}{\d u^k} \, 
    \left. \Big( \Tilde{\beta}_s^{\varepsilon, \eta}(u) \,
    \i \, \mL_{ \frac{\d}{\d s}\Tilde{S}_s^{\varepsilon, \eta}}
    \, \Tilde{\beta}_s^{\varepsilon, \eta}(-u) \, B \Big) \right|_{u=0}
    \\
    &\quad +  \int_0^1 \d \lambda \, \frac{\lambda^{n+1}}{(n + 1) !} \,  \frac{\d^{n+1}} {\d u^{n + 1}} 
    \, \left. \left( \Tilde{\beta}_s^{\varepsilon,\eta}(u)
    \, \i \, \mL_{ \frac{\d}{\d s}\Tilde{S}_s^{\varepsilon, \eta}}
    \, \Tilde{\beta}_s^{\varepsilon, \eta} (-u) \, B \right) \right|_{u = \tau(\lambda)}
    \\
    & = \sum_{k = 0}^n \frac{1}{(k + 1)!} \, \i \, \mL_{\ad(\i \Tilde{S}_s^{\varepsilon, \eta} )^k \, \frac{\d}{\d s} \Tilde{S}_s^{\varepsilon, \eta}} \, B 
    \\
    &\quad + \int_0^1 \d \lambda \, \frac{\lambda^{n + 1}}{(n + 1)!} \,  
    \,  \left( \Tilde{\beta}_s^{\varepsilon,\eta}(\tau(\lambda))
    \, \i \, \mL_{ \ad( \i \Tilde{S}_s^{\varepsilon, \eta})^{n + 1}  \frac{\d}{\d s}\Tilde{S}_s^{\varepsilon, \eta}}
    \, \Tilde{\beta}_s^{\varepsilon, \eta} (-\tau(\lambda)) \, B \right)    
    \, ,
\end{align*}
where $\tau(\lambda) \in [0,1]$ is a constant dependent on $\lambda$ and $\ad(\cdot)^k$ denotes the $k$-times iterated commutator.
For the third term, we proceed similarly:
\begin{align*}
    \hspace{2em}&\hspace{-2em} \left. \Tilde{\beta}_s^{\varepsilon, \eta} ( \mu)  \, \i\, \mL_{ H_s^{\varepsilon, \eta} } \, \Tilde{\beta}_s^{\varepsilon, \eta} ( - \mu) \, B \, \right|_{\mu = 1}
    \\
    &= \left. \sum_{k=0}^n  \frac{1}{k!} \, \frac{\d^k}{\d u^k} \, \left( \Tilde{\beta}_s^{\varepsilon, \eta}(u) \, \i \, \mL_{H_s^{\varepsilon, \eta}} \, \Tilde{\beta}_s^{\varepsilon, \eta}(-u) \, B \right) \right|_{u=0}
    \\
    &\quad + \left. \frac{1}{(n + 1) !}  \frac{\d^{n+1}} {\d u^{n + 1}} \, \left( \Tilde{\beta}_s^{\varepsilon,\eta} (u) \, \i \, \mL_{H_s^{\varepsilon, \eta}} \, \Tilde{\beta}_s^{\varepsilon, \eta} (-u) \, B \right) \right|_{u = \tau}
    \\
    & = \sum_{k=0}^n \frac{1}{k!}\, \, \i \,\mL_{\ad(\i \Tilde{S}_s^{\varepsilon,\eta})^k H_s^{\varepsilon,\eta}} \, B
    \\
    & \quad + \frac{1}{( n + 1) !} \, \Tilde{\beta}_s^{\varepsilon,\eta} (\tau) \, \i \, \mL_{\ad( \i \Tilde{S}_s^{\varepsilon, \eta})^{n + 1} H_s^{\varepsilon, \eta}} \, \Tilde{\beta}_s^{\varepsilon,\eta} (-\tau) \, B \, ,
\end{align*}
where $\tau \in [0,1]$. 

Combining these two Taylor expansions and the term we get from the derivative of the ground state, we have in total
\begin{align*}\label{eq: terms after taylor}
    \hspace{4ex} & \hspace{-4ex} \frac{\d}{\d s} \, \w_s \Big( \Tilde{\beta}_s^{\varepsilon, \eta}(1) \, \mU_{s, t}^{\varepsilon, \eta} \, A \Big) 
    \\
    &=  -\, \w_s \left( \sum_{k = 0}^n \i \, \mL_{ \frac{1}{k!} \, \ad( \i \Tilde{S}_s^{\varepsilon,\eta})^k H_s^{\varepsilon,\eta} - \frac{1}{(k + 1)!} \, \ad(\i \Tilde{S}_s^{\varepsilon, \eta} )^k \, \frac{\d}{\d s} \Tilde{S}_s^{\varepsilon, \eta} +  \mI_s( \Dot{H}_s )} \, B \right) 
    \\
    &\quad +  \w_s \left( \int_0^1 \d \lambda \, \frac{\lambda^{n + 1}}{(n + 1)!} \,    
    \,  \left( \Tilde{\beta}_s^{\varepsilon,\eta}(\Tilde{\tau}(\lambda))
    \, \i \, \mL_{ \ad( \i \Tilde{S}_s^{\varepsilon, \eta})^{n + 1}  \frac{\d}{\d s}\Tilde{S}_s^{\varepsilon, \eta}}
    \, \Tilde{\beta}_s^{\varepsilon, \eta} (-\Tilde{\tau}(\lambda)) \, B \right)  \right) \numberthis
    \\
    & \quad -  \w_s \left( \frac{1}{( n + 1) !} \, \Tilde{\beta}_s^{\varepsilon,\eta} (\tau) \, \i \, \mL_{\ad( \i \Tilde{S}_s^{\varepsilon, \eta})^{n + 1} H_s^{\varepsilon, \eta}} \, \Tilde{\beta}_s^{\varepsilon,\eta} (-\tau) \, B \right) \, .
\end{align*}
The two remainder terms of the Taylor expansions will enter in the final bound. 
We continue by expanding the first term in powers of $\eps$ and $\eta$ and then make a concrete choice of the interactions $ K^{j,i}_t $ that will make this expansion vanish up to a third remainder term. The interaction appearing in the first term is given by
\begin{align*}\label{eq: expanison}
    \hspace{2em} & \hspace{-2em} \sum_{k = 0}^n \frac{1}{k!} \, \ad( \i \Tilde{S}_s^{\varepsilon,\eta})^k H_s^{\varepsilon,\eta} 
    - \sum_{k = 0}^n \frac{1}{(k + 1)!} \, \ad(\i \Tilde{S}_s^{\varepsilon, \eta} )^k \, \frac{\d}{\d s} \Tilde{S}_s^{\varepsilon, \eta} 
    +  \mI_s( \Dot{H}_s ) \numberthis
    \\
    & = \frac{1}{\eta} \, \left( \sum_{k = 0}^n \frac{1}{k !} \, \ad( \i \, \Tilde{S}_s^{\varepsilon, \eta})^k \left( H_s + \eps \, H_s^1 + \eps\, V_s \right) - \sum_{k = 0}^n \frac{1}{(k + 1)!} \, \eta \, \ad( \i \, \Tilde{S}_s^{\varepsilon, \eta})^k \, \frac{\d}{\d s} \Tilde{S}_s^{\varepsilon, \eta} + \eta \, \mI_s (\Dot{H}_s) \right)
    \, .
\end{align*}
We look at the factor without the inverse power of $\eta$ and define
\begin{align*}
    \hspace{4em}&\hspace{-4em} \sum_{k = 0}^n \frac{1}{k !} \, \ad( \i \, \Tilde{S}_s^{\varepsilon, \eta})^k \left( H_s + \eps \, H_s^1 + \eps\, V_s \right) - \sum_{k = 0}^n \frac{1}{(k + 1)!} \, \eta \, \ad( \i \, \Tilde{S}_s^{\varepsilon, \eta})^k \, \frac{\d}{\d s} \Tilde{S}_s^{\varepsilon, \eta} + \eta \, \mI_s (\Dot{H}_s) 
    \\
    &\eqcolon  H_s + \sum_{j=1}^n \sum_{i=0}^j \varepsilon^i \eta^{j - i} C_s^{j, i} +  \sum_{i=0}^{n+1} \varepsilon^{i} \eta^{n + 1 -i} R_s^{i}(\eps,\eta)  \, , 
\end{align*}
with $C_s^{j,i}$ being a sum of $B_\infty$-interactions and Lipschitz potentials and $R_s^{i}(\eps,\eta)$ polynomials in $\eps$ and $\eta$ with coefficients in $B_\infty$.

Plugging the ansatz  \eqref{def: S^n} for $\Tilde{S}_s^{\varepsilon,\eta}$ into this expression we find  
\begin{align*}
    C_s^{1,1} &=  \i \,  [K_s^{1,1}, \, H_s] + (H_s^1 + V_s)
    \\
    C_s^{1,0} &=   \i \, [K_s^{1,0}, \, H_s] + \mI_s( \Dot{H}_s)
    \\
    C_s^{2,2} &=   \i \, [K_s^{2,2}, \, H_s] - \frac{1}{2} [K_s^{1,1}, \, [K_s^{1,1}, \, H_s]] + \i [K_s^{1,1}, \, (H_s^1 + V_s)]
    \\
    &\mathrel{} \vdots
\end{align*}
We observe that, in general, the interactions $C_s^{j, i}$ can always be written as
\begin{align*}
    C_s^{j,i} =  \i \, [K_s^{j,i}, \, H_s] + L_s^{j, i} \, ,
\end{align*}
where the interaction $L_s^{j,i}$ is given as by sums and iterated commutators of $H_s$, $(H_s^1 + V_s)$, $\mI_s(\Dot{H}_s)$ and $K_s^{l, m}$ for $l < j$ and derivatives thereof. This tells us that for each $ j \leq n $  we can iteratively choose the interactions $K_s^{j,i}$ as $K_s^{j,i} = - \mI_s( L_s^{j,i})$. This iterative choice results in a cancellation of most of the terms, since by Lemma \ref{lem: liouvillian of inverse liouvillian}, we have $C_s^{j,i} = - (L_s^{j,i})\OD[s] + L_s^{j,i}$ and
\begin{align*}\label{eq: cancelation}
    & \tfrac{-1}{\eta}\, \w_s( \i \, \mL_{ H_s + \sum_{j=1}^n \sum_{i=0}^j \varepsilon^i \eta^{j - i} C_s^{j,i} + \sum_{i = 0}^{n + 1} \varepsilon^{i} \eta^{n + 1 - i} R_s^{i}(\eps,\eta)} \, B)
    = \tfrac{-1}{\eta}\, \w_s( \i \, \mL_{\sum_{i = 0}^{n + 1} \varepsilon^{i} \eta^{n + 1 - i} R_s^{i}(\eps,\eta)}\, B) \, , \numberthis
\end{align*}
where the first term vanishes, since $\w_s$ is an $\mL_{H_s}$ ground state. 

Furthermore, note that our choice also ensures that $(K_s^{j,i})_{s \in I} \in B_{\infty, I, L}^{(\infty)}$ due to Lemma \ref{lem: comm is smooth} and Lemma \ref{lem: inverse liouvillian of b_infty} as claimed above. Moreover, by Lemma \ref{lem: inverse liouvillian of b_infty} it follows that
\begin{align*}\label{eq: locality of K}
    \sup_{s \in I} \, \sup_{x \in \Z^d} \,  \lVert (K_s^{j,i})_x \rVert_{\nu, x} \, (1+ \d (x, L))^m < \infty \numberthis
\end{align*}
for all $j, i \leq n$ and $\nu, m \in \N_0$. We will use this observation, when proving the locality statement for $S_t^{\eps, \eta}$.

\subsection{Bounding the Remainder Terms}

At this point, there are three remainder terms, one that is left after  the cancellations in \eqref{eq: cancelation} and two from \eqref{eq: terms after taylor}, resulting from the two Taylor expansions. We need to show that each of them satisfies the bound \eqref{eq:n-bound}. We begin with the one from \eqref{eq: cancelation}
\begin{align*} \label{eq: bound taylor expansion}
        \rvert \tfrac{1}{\eta}\, \w_s( \i \, \mL_{\sum_{i = 0}^{n + 1} \varepsilon^{i} \eta^{n + 1 - i} R_s^{i}(\eps,\eta)}\, B) \rvert 
        \leq \tfrac{1}{\eta}\, \sum_{i=0}^{n+1} \max\left(\eps,\eta\right)^{n+1} \,  \lvert \w_s(  \i \, \mL_{R_s^{i}(\eps,\eta)} \, \tilde{\beta}_{s}^{\eps, \eta}(1)\, \mathfrak{U}_{s,t}^{\eps,\eta} \, A) \rvert . \numberthis
\end{align*}
We apply Lemma \ref{Lemma: final bound} to each of the summands and obtain the bound
\begin{align*}
    \lvert \w_s(  \i \, \mL_{R_s^{i}(\eps,\eta)} \, \tilde{\beta}_{s}^{\eps, \eta}(1)\, \mathfrak{U}_{s,t}^{\eps,\eta} \, A) \rvert &\leq \lVert \mL_{R_s^{i}(\eps,\eta)} \, \tilde{\beta}_{s}^{\eps, \eta}(1)\, \mathfrak{U}_{s,t}^{\eps,\eta} \, A \rVert
    \\
    &\leq c_i \left( 1 +  \left( \tfrac{|t - s|}{\eta} \right)^{\kappa d} \, f \left( \tfrac{\lvert t - s \rvert}{\eta} \right) \right) \, \lVert A \rVert \, |M|^2 \, ,
\end{align*}
where we have included the factor $\lVert R_s^i (\eps, \eta) \rVert_{d + 1} $ in the constant $n$-dependent constant $c_i$, since it can be bounded above uniformly in $s$, $\eps$, and $\eta$.
Plugging this into equation \eqref{eq: bound taylor expansion} we find
\begin{align*} 
    \left| \tfrac{1}{\eta}\, \w_s( \i \, \mL_{\sum_{i = 0}^{n + 1} \varepsilon^{i} \eta^{n + 1 - i} R_s^{i}(\eps,\eta)}\, B) \right| 
    & \leq \max \left(\eps, \eta \right)^{n+1} \, \tfrac{1}{\eta} \, \sum_{i = 0}^{n + 1} c_i \left( 1 + \left( \tfrac{|t - s|}{\eta} \right)^{\kappa d} \, f \left( \tfrac{\lvert t - s \rvert}{\eta} \right) \right) \, \lVert A \rVert \, |M|^2 \,
    \\
    & \leq \Tilde{c}_n \, \max \left(\eps, \eta \right)^{n+1} \, \tfrac{1}{\eta}  \, \left( 1 +  \left(\tfrac{|t - t_0|}{\eta}\right)^{\kappa d} \, f \left( \tfrac{\lvert t - t_0 \rvert}{\eta} \right) \right) \, \lVert A \rVert \,  |M|^2 \, ,
\end{align*}
where we have used that $s \in [t_0, t]$. Next, we show that the same bound also holds for the remainder terms from the Taylor expansions \eqref{eq: terms after taylor}. We start with
\begin{align*}
    \w_s \left( \int_0^1 \d \lambda \, \frac{\lambda^{n + 1}}{(n + 1)!} \, 
    \, \left( \Tilde{\beta}_s^{\varepsilon,\eta}(\Tilde{\tau}(\lambda))
    \, \i \, \mL_{ \ad( \i \Tilde{S}_s^{\varepsilon, \eta})^{n + 1}  \frac{\d}{\d s}\Tilde{S}_s^{\varepsilon, \eta}}
    \, \Tilde{\beta}_s^{\varepsilon, \eta} (1 - \Tilde{\tau}(\lambda)) \, \mU_{s, t}^{\varepsilon, \eta} \, A  \right)   \right)
    \, .
\end{align*}
Since each term of the sum $\Tilde{S}_s^{\eps, \eta}$ contains at least one $\eps$ or $\eta$, the interaction appearing in this term is at least of order $n + 2$ in $\eps$ and $\eta$. We can therefore write the interaction as 
\begin{align*}
    \ad( \i \Tilde{S}_s^{\varepsilon, \eta})^{n + 1}  \frac{\d}{\d s}\Tilde{S}_s^{\varepsilon, \eta}
    \eqcolon \sum_{i=0}^{n+2} \eps^i \, \eta^{n+2-i} \, R_{s,2}^i(\eps,\eta) 
    \, ,
\end{align*}
with $R_{s,2}^i(\eps,\eta)$ being a polynomial in $\eps$ and $\eta$ with coefficients in $B_{\infty}$. We find
\begin{align*}
    \hspace{12ex} & \hspace{-12ex} \left| \w_s \left( \int_0^1 \d \lambda \, \frac{\lambda^{n + 1}}{(n + 1)!} \,   
    \, \left( \Tilde{\beta}_s^{\varepsilon,\eta}(\Tilde{\tau}(\lambda))
    \, \i \, \mL_{ \ad( \i \Tilde{S}_s^{\varepsilon, \eta})^{n + 1}  \frac{\d}{\d s}\Tilde{S}_s^{\varepsilon, \eta}}
    \, \Tilde{\beta}_s^{\varepsilon, \eta} (1 - \Tilde{\tau}(\lambda)) \, \mU_{s, t}^{\varepsilon, \eta} \, A  \right) \right) \right| 
    \\
    &  \leq \max \left( \eps, \eta \right)^{n + 2} \int_0^1 \d \lambda \, \frac{\lambda^{n + 1}}{(n + 1)!} \, \sum_{i = 0}^{n + 2} \, \lVert \mL_{R_{s, 2}^i(\eps, \eta)} \, \Tilde{\beta}_s^{\eps, \eta} (1 - \Tilde{\tau}(\lambda)) \, \mU_{s, t}^{\eps, \eta} \, A \rVert \, .
\end{align*}
We now proceed as before and bound each summand. To this end, we again use Lemma \ref{Lemma: final bound} and find
\begin{align*}
    \lVert \mL_{R_{s, 2}^i (\eps, \eta)} \, \Tilde{\beta}_s^{\eps, \eta} (1- \Tilde{\tau}(\lambda)) \, \mU_{s,t}^{\eps, \eta} \, A \rVert \leq c_i \left( 1 + \left( \tfrac{|t - s|}{\eta} \right)^{\kappa d}  \, f \left( \tfrac{\lvert t - s \rvert}{\eta} \right) \right) \, \lVert A \rVert \, |M|^2 \, ,
\end{align*}
where we have put the dependence on $R_{s,2}^i(\eps, \eta)$ into the constant $c_i$, since this expression can be bounded uniformly in $\eps$, $\eta$ and $s \in I$. Note that this bound is independent of $\Tilde{\tau}(\lambda)$. We therefore have
\begin{align*}
    \hspace{24ex} & \hspace{-24ex} \left| \w_s \left( \int_0^1 \d \lambda \, \frac{\lambda^{n + 1}}{(n + 1)!} \, 
    \, \left( \Tilde{\beta}_s^{\varepsilon,\eta}(\Tilde{\tau}(\lambda))
    \, \i \, \mL_{ \ad( \i \Tilde{S}_s^{\varepsilon, \eta})^{n + 1}  \frac{\d}{\d s}\Tilde{S}_s^{\varepsilon, \eta}}
    \, \Tilde{\beta}_s^{\varepsilon, \eta} (1 - \Tilde{\tau}(\lambda)) \, \mU_{s, t}^{\varepsilon, \eta} \, A  \right)   \right) \right| 
    \\
    &\leq c^\prime_n \, \max \left( \eps, \eta \right)^{n + 2} \, \left( 1 + \left( \tfrac{|t - t_0|}{\eta} \right)^{\kappa d} \, f \left( \tfrac{\lvert t - t_0 \rvert}{\eta} \right) \right) \, \lVert A \rVert \, |M|^2 \, ,
\end{align*}
where we have again used that $s \in [t_0, t]$. This bound is now in the same form as the bound for the first term and the third remainder term can be handled in exactly the same way. Hence, combining the three remainder terms results in
\begin{align*}
    \left| \frac{\d}{\d s} \, \w_s \left( \Tilde{\beta}_s^{\varepsilon, \eta} (1) \, \mU_{s, t}^{\varepsilon, \eta} \, A \right) \right| 
    \leq   c_n \, \max \left( \varepsilon, \eta \right)^{n + 1} \, \tfrac{1}{\eta} \, \left( 1 + \left( \tfrac{|t - t_0|}{\eta} \right)^{\kappa d} \, f \left( \tfrac{\lvert t - t_0 \rvert}{\eta} \right) \right) \, \lVert A \rVert \, |M|^2 \, ,
\end{align*}
from which it immediately follows that 
\begin{align*} \label{equation: bound}
    |\Tilde{\w}_t^{\varepsilon,\eta}(A) - \Tilde{\w}_{t_0}^{\varepsilon, \eta} (\mU_{t_0, t}^{\varepsilon, \eta} \, A)| & \leq \int_{t_0}^t \d s \, \left| \frac{\d}{\d s} \, \w_s \left( \Tilde{\beta}_s^{\varepsilon, \eta} (1) \, \mU_{s, t}^{\varepsilon, \eta} \, A \right) \right|
    \\
    &\leq c_n \, \max \left( \varepsilon, \eta \right)^{n + 1} \, \tfrac{|t - t_0|}{\eta} \left( 1 + \left( \tfrac{|t - t_0|}{\eta} \right)^{\kappa d} \, f \left( \tfrac{\lvert t - t_0 \rvert}{\eta} \right) \right) \, \lVert A \rVert \, |M|^2 \, . \numberthis
\end{align*}
This shows the bound for the $n$-dependent super-adiabatic approximation.

\subsection{Construction of the $n$-independent generator}

We now come to the second part of the proof, in which we construct an $n$-independent family of interactions $(S_t^{\eps,\eta})_{t\in I}$ from the $n$-dependent family of the first part by a suitable resummation. From our construction it then follows directly that three key properties hold: the $n$-independent states $\w_t^{\eps, \eta} \coloneq \w_t \circ \exp( \i \mL_{S_t^{\eps, \eta}})$ satisfy the same bound \eqref{equation: bound} as the $n$-dependent states, the quantity $\sup_{(\eps, \eta) \in [0,1]^2} \, \sup_{s \in I} \, \lVert S_s^{\varepsilon, \eta} \rVert_{\nu}$ is finite for all $\nu \in \N_0$ and $S_s^{\eps, \eta}$ is localized in $L$. 

We begin by setting $\chi \coloneq \chi_{[0, \, 1]}$, where $\chi_{[0, \, 1]}$ denotes the characteristic function of the interval $[0, \, 1]$ and define
\begin{align} \label{def: S}
    S_s^{\varepsilon, \eta} \coloneq \sum_{j = 1}^\infty \chi \left( \varepsilon / \delta_j \right) \, \chi \left( \eta / \delta_j \right) \sum_{i=0}^{j} \varepsilon^i \, \eta^{j-i} \, K_s^{j,i} \, ,
\end{align}
where $(\delta_j)_{j \in \N}$ is a sequence that is non-zero, monotonically decreasing, and converging to zero. It is defined by $\delta_1 \coloneq 1$ and 
\begin{align*}
    \delta_j \coloneq \min \left(\frac{1}{2^j}, \, \delta_1 , \, ... \, , \delta_{j-1}, \,  \frac{1}{f_j}, \, \frac{1}{g_j}, \, \frac{1}{h_j} \right) \, ,
\end{align*}
for $j > 1$. The first elements of the $\min$ ensure that the sequence converges monotonically to $0$, while the three sequences $(f_j)_{j\in \N}$, $(g_j)_{j\in \N}$ and $(h_j)_{j\in \N}$ ensure that the convergence is fast enough that the three key properties hold. They are defined as
\begin{align*}
    f_j & \coloneq 2 \max\left(1, \max_{\substack{1 \leq l \leq j}} c_l  \right)
    \\
    g_j & \coloneq 2 \max\left(1, \max_{\substack{1 \leq l \leq j\\ 0\leq m\leq j}} \left(\sup_{s\in I} \lVert K_s^{l,m} \rVert_j \right)\right)
    \\
    h_j & \coloneq 2 \max\left(1, \max_{\substack{1 \leq l \leq j\\ 0\leq m\leq j}} \left(\sup_{x \in \Z^d} \sup_{s \in I} \lVert (K_s^{l, m})_x \rVert_{j, x} (1 + \d (x, L))^j \right)\right) 
    \, ,
\end{align*}
where $c_l$ are the constants from bound  \eqref{equation: bound} that we obtained in the first part.

Note that terms in the sum \eqref{def: S} proportional to $\varepsilon^l \, (\frac{\eta}{\varepsilon})^m$, with $m \leq l$ and $l \leq j$, are included if and only if $\max \left( \varepsilon, \eta \right) \leq \delta_j$. Hence, the sum is finite for every fixed $\eps$ and $\eta$. Since every summand lies in $B_{\infty, I, L}^{(\infty)}$, we have $(S_s^{\varepsilon, \eta} )_{s \in I} \in B_{\infty, I, L}^{(\infty)}$.

To prove that the bound claimed in the theorem holds in the $n$-independent case, we define the constant
\begin{align*}
  C_n \coloneq 
  \max \left(1, \, \max_{0 \leq k < n} \left( \frac{ c_k \delta_k^{k + 1} }{ \delta_{k + 1}^{n + 1}} \right) \right ) 
  \, ,  
\end{align*}
where $c_k$ is again the constant obtained in the bound of Equation \eqref{equation: bound}. For this discussion, we choose $k \in \N$ such that $\delta_{k+1} < \max \left( \varepsilon, \eta \right) \leq \delta_k$. This choice is always possible by the definition of the sequence.  We split the argument into two cases. First, for $n + 1 \leq k$, we find
\begin{align*}
    c_k \max \left( \varepsilon, \eta \right)^{k+1} &\leq c_k \max \left( \varepsilon, \eta \right) \, \max \left(\varepsilon, \eta \right)^{n+1} \leq c_k \, \delta_k \, \max \left(\varepsilon, \eta \right)^{n + 1}
    \\
    &\leq \max \left( \varepsilon, \eta\right)^{n + 1} \leq C_n \max \left( \varepsilon, \eta \right)^{n + 1} \, .
\end{align*}
Similarly, for $n + 1 > k$ we find
\begin{align*}
    c_k \max \left( \varepsilon, \eta\right)^{k + 1} &\leq \frac{c_k \delta_k^{k + 1}}{\max \left( \varepsilon, \eta\right)^{n + 1}} \max \left( \varepsilon, \eta \right)^{n + 1} \leq \frac{c_k \delta_k^{k + 1}}{\delta_{k + 1}^{n + 1}} \max \left( \varepsilon, \eta \right)^{n + 1}
    \\
    &\leq C_n \max \left( \varepsilon, \eta \right)^{n + 1} \, .
\end{align*}
We now use that $\w_t^{\varepsilon, \eta}$ is equal to the $n$-dependent state $\tilde{\w}_t^{\varepsilon, \eta}$ for $n=k$ and combine the above statements with the bound $\eqref{equation: bound}$ and get for all $n\in \N$ that
\begin{align*}
    \hspace{4ex} \hspace{-4ex} |\w_t^{\varepsilon, \eta} (A) - \w_{t_0}^{\varepsilon, \eta}  (\mU_{t_0, t}^{\varepsilon, \eta} \, A )|
    &= |\Tilde{\w}_t^{\varepsilon, \eta} (A) - \Tilde{\w}_{t_0}^{\varepsilon, \eta}  (\mU_{t_0, t}^{\varepsilon, \eta} \, A) |
    \\
    &\leq c_k  \max \left( \varepsilon, \eta\right)^{k + 1} \, \tfrac{|t - t_0|}{\eta} \, \left( 1 + \left( \tfrac{|t - t_0|}{\eta} \right)^{\kappa d} \, f \left( \tfrac{\lvert t - t_0 \rvert}{\eta} \right) \right) \, \lVert A \rVert \, |M|^2
    \\
    &\leq C_n  \max \left( \varepsilon, \eta \right)^{n + 1} \, \tfrac{|t - t_0|}{\eta} \, \left( 1 + \left( \tfrac{|t - t_0|}{\eta} \right)^{\kappa d} \, f \left( \tfrac{\lvert t - t_0 \rvert}{\eta} \right) \right) \, \lVert A \rVert \, |M|^2
    \\
    &\leq C_n \left( \varepsilon^{n + 1} + \eta^{n + 1} \right) \, \tfrac{|t - t_0|}{\eta} \, \left( 1 + \left( \tfrac{|t - t_0|}{\eta} \right)^{\kappa d} \, f \left( \tfrac{\lvert t - t_0 \rvert}{\eta} \right) \right) \, \lVert A \rVert \, |M|^2 \, ,
\end{align*}
which is the desired bound. At this point, note that in the case where $A$ is not strictly localized but $A \in \mA_\infty$, Lemma \ref{Lemma: D_infty case} says that we get the bound
\begin{align*}
    \hspace{4ex} \hspace{-4ex} |\w_t^{\varepsilon, \eta} (A) - \w_{t_0}^{\varepsilon, \eta}  (\mU_{t_0, t}^{\varepsilon, \eta} \, A )| 
    \leq \tilde{C}_n \left( \varepsilon^{n + 1} + \eta^{n + 1} \right) \, \tfrac{|t - t_0|}{\eta} \, \left( 1 + \left( \tfrac{|t - t_0|}{\eta} \right)^{\kappa d} \, f \left( \tfrac{\lvert t - t_0 \rvert}{\eta} \right) \right) \, \lVert A \rVert_{2d+2} 
    \, ,
\end{align*}
with a different constant $\tilde{C}_n$. The conditions of this lemma are satisfied due to Lemmas~\ref{lem: limit in decay norm}, \ref{lem: sum representation of generator}, and \ref{lem: cocycles}.

Now, we will prove that for any $\nu \in \N_0$ the interaction norm $\lVert S_s^{\eps, \eta} \rVert_{\nu}$ is uniformly bounded in $s 
$, $\eps$, and $\eta$. Let again $k \in \N$ such that $\delta_{k + 1} < \max \left( \eps, \, \eta \right) \leq \delta_k$. We divide the proof into two cases. First, for $k \geq \nu$, we have for all $s \in I$
\begin{align*}
    \lVert S_s^{\varepsilon,\eta} \rVert_\nu \leq & \sum_{j = 1}^k \sum_{i = 0}^j \varepsilon^{i} \eta^{j - i} \lVert K_s^{j,i} \rVert_\nu \leq \sum_{j = 1}^k \sum_{i=0}^j \delta_k^{j} \lVert K_s^{j,i} \rVert_\nu
    \\
    \leq & \sum_{j = 1}^k (j + 1) \, 2^{- j} \leq 3 \, .
\end{align*}
Now, consider the case $k < \nu$. We find
\begin{align*}
    \lVert S_s^{\varepsilon, \eta} \rVert_\nu \leq & \sum_{j=1}^k \sum_{i=0}^j \varepsilon^{i} \eta^{j - i} \lVert K_s^{j, i} \rVert_\nu \leq \sum_{j = 1}^\nu \sum_{i = 0}^j \lVert K_s^{j, i} \rVert_\nu \, .
\end{align*}
This expression is independent of $\eps$, $\eta$, and $k$. Combining the two cases yields for all $\nu \in \N_0$ the statement
\begin{align*}
    \sup_{(\varepsilon, \eta) \in [0,1]^2 } \, \sup_{s \in I} \, \lVert S_s^{\varepsilon, \eta} \rVert_\nu \leq \max \left( 3, \, \sum_{j = 1}^\nu \sum_{i = 0}^j \,  \sup_{s \in I} \, \lVert K_s^{j, i} \rVert_\nu \right) \, < \infty \, .
\end{align*}
The localization of $(S_s^{\eps, \eta})_{s \in I}$ in the set $L$ follows by the same calculation when replacing the interaction norm $\lVert \, \cdot \, \rVert_\nu$ by $\sup_{x \in \Z^d}  \lVert ( \, \cdot \,  )_x \rVert_{\nu, x} (1 + \d (x, L))^m$ and one obtains for all $\nu, m \in \N_0$ that
\begin{align*}
    \hspace{2em}&\hspace{-2em}\sup_{(\varepsilon, \eta) \in [0,1]^2 } \, \sup_{s \in I} \, \sup_{x \in \Z^d}  \, \lVert ( S_s^{\eps, \eta}  )_x \rVert_{\nu, x} \, (1 + \d (x, L))^m 
    \\
    &\leq \max \left( 3, \, \sum_{j = 1}^{\max(\nu,m)} \sum_{i = 0}^j \,  \sup_{s \in I} \sup_{x \in \Z^d}  \lVert (K_s^{j, i}  )_x \rVert_{\nu, x} \,  (1 + \d (x, L))^m\, \right)
    < \infty 
    \, ,
\end{align*}
which is finite due to Statement \eqref{eq: locality of K}.

\subsection{Additional Properties of $S_t^{\eps,\eta}$}

In this section, we discuss the additional properties of the super-adiabatic approximation claimed in the main theorem.

\paragraph{Locality of $S_t^{\epsi,\eta}$ in time and space:}
The locality in time of $S_t^{\eps, \eta}$ follows directly from its construction ($K_t^{j,i}$ is chosen in a way that depends only on $H_t$, $H^1_t$, $V_t $ and their time derivatives at time $t$.) and we have discussed the locality in space in the previous section.

\paragraph{Stationarity:} In the case, where the Hamiltonian $H_t^{\varepsilon,\eta}\equiv H^{\epsi,\eta}_J$ is constant on some interval $J\subseteq I$, using the locality in time of $K_t^{j,i}$, we find that the second and third expression in Equation \eqref{eq: expanison} vanish. It follows that in the iterative construction only interaction terms $K_t^{j, i}$ with $i = j$ are non-zero. Following the proof of the main theorem with the simplified expression, we arrive at the bound claimed for the NEASS ($\max \left( \eps, \eta \right)$ is replaced with $\eps$ and $\eta$ gets set to $1$ in all other places).

\paragraph{Vanishing Perturbation:} In the case where for some $t \in I$ all time derivatives of $H_t^{\eps, \eta}$ vanish and $H_t^1=V_t=0$, the construction yields $K_t^{j,i} = 0$ for all $i$ and $j$, which implies $\w_{t}^{\eps,\eta}= \w_t$.

\appendix

\section{Bound for the Remainder Terms}

\medskip
\begin{lemma} \label{Lemma: final bound}
     Let $R \in B_\infty$, $A \in \mA_X$ and $\Tilde{\beta}_s^{\varepsilon,\eta} (u)$, $\mU_{s, t}^{\varepsilon, \eta}$ as in Section \ref{Section: Proof} with $0 \leq u \leq 1$. It holds that
    \begin{align*}
        \lVert \mL_{R} \, \Tilde{\beta}_s^{\varepsilon,\eta} (u) \, \mU_{s, t}^{\varepsilon, \eta} \, A \rVert \leq  C \, \lVert R \rVert_{d + 1} \, \lVert A \rVert \, |X|^2 \left( 1 + \left( \tfrac{|t - s|}{\eta} \right)^{\kappa d} \, f \left( \tfrac{1}{\eta} \, \lvert t - s\rvert \right) \right)  \, ,
    \end{align*}
    where $C$ is some constant independent of $t$, $s$, $u$, $\eta$, $\eps$ and $A$. Here, $f$ is the function and $\kappa$ the constant from Assumption \hyperref[ass: LR bound]{(\ref*{ass: LR bound})}.
\end{lemma}
\begin{proof}
    Let $c >0$ and $X_c$ be the fattening of $X$ by $c$ i.e. 
    $$X_c\coloneq \{ y \in \Z^d \,\, | \,\, \dd(y,X) \leq c \}\, .$$ 
    We can use Lemma \ref{lem: sum representation of generator} and split the sum into three parts
    \begin{align*}
        \lVert \mL_{R} \, \Tilde{\beta}_s^{\varepsilon, \eta} (u) \, \mU_{s, t}^{\eps, \eta} \, A \rVert
        &\leq \sum_{x\in \Z^d} \lVert [\, \Tilde{\beta}_s^{\varepsilon, \eta} (-u) \, R_x ,\, \mU_{s, t}^{\eps, \eta} \, A \,] \rVert
        \\
        &\leq  \sum_{x\in X_c} \lVert [\, \Tilde{\beta}_s^{\varepsilon, \eta} (-u) \, R_x ,\, \mU_{s, t}^{\eps, \eta} \, A \,] \rVert
        \\
        &\quad+ \sum_{x\in \Z^d \setminus X_c} \lVert [\, (1- \E_{B_{\frac{\dd(x,X)}{2}-\frac{c}{2}}(x)})\, \Tilde{\beta}_s^{\varepsilon, \eta} (-u)\, R_x ,\, \mU_{s, t}^{\eps, \eta} \, A \,] \rVert
        \\
        &\quad+ \sum_{x\in \Z^d \setminus X_c} \lVert [\, \E_{B_{\frac{\dd(x,X)}{2}-\frac{c}{2}}(x)} \, \Tilde{\beta}_s^{\varepsilon, \eta} (-u) \, R_x ,\, \mU_{s, t}^{\eps, \eta} \, A \,] \rVert
    \end{align*}
    We proceed by bounding each part separately. For the first sum we use the simple bound
    \begin{align*}
        \sum_{x\in X_c} \lVert [\, \Tilde{\beta}_s^{\varepsilon, \eta} (-u) \, R_x ,\, \mU_{s, t}^{\eps, \eta} \, A \,] \rVert
        \leq 2\, \lvert X_c \rvert \, \,\lVert R \rVert_0\, \lVert A \rVert\, .
    \end{align*}
    We move on to the second sum, which we estimate by
    \begin{align*}
        \hspace{2em}&\hspace{-2em} \sum_{x\in \Z^d \setminus X_c} \lVert [\, (1- \E_{B_{\frac{\dd(x, X)}{2}-\frac{c}{2}}(x)})\, \Tilde{\beta}_s^{\varepsilon, \eta} (-u)\, R_x ,\, \mU_{s, t}^{\eps, \eta} \, A \,] \rVert
        \\
        &\leq \sum_{x\in \Z^d \setminus X_c} 2\, \lVert (1- \E_{B_{\frac{\dd(x,X)}{2}-\frac{c}{2}}(x)})\, \Tilde{\beta}_s^{\varepsilon, \eta} (-u) \, R_x \rVert \, \lVert A \rVert
        \\
        &\leq \sum_{y\in X}\sum_{x\in \Z^d \setminus X_c} 2 \, \lVert (1- \E_{B_{\frac{\lVert x-y \rVert}{2}-\frac{c}{2}}(x)})\, \Tilde{\beta}_s^{\varepsilon, \eta} (-u) \, R_x \rVert \, \lVert A \rVert
        \\
        &\leq \sum_{y\in X} \sum_{k= \lceil c \rceil }^\infty 2d\, (2k)^{d-1} \,  2 \, \sup_{x \in \Z^d} \, \lVert (1- \E_{B_{\frac{k}{2}-\frac{c}{2}}(x)})\, \Tilde{\beta}_s^{\varepsilon, \eta} (-u) \, R_x \rVert \, \lVert A \rVert
        \\
        &\leq \sum_{y\in X} \sum_{k= \lceil c \rceil}^\infty 2d\, (2k)^{d-1} \,  2 \, \sup_{x \in \Z^d} \frac{\lVert \Tilde{\beta}_s^{\varepsilon, \eta} (-u) \, R_x  \rVert_{d+1, x}}{(1 + \frac{k}{2}-\frac{c}{2})^{d+1}}\,  \lVert A \rVert
        \\
        &\leq \sum_{y\in X} \sum_{k= 0}^\infty 2d\, (2k+ 2\lceil c \rceil)^{d-1} \,  2 \, \sup_{x \in \Z^d} \frac{\lVert \Tilde{\beta}_s^{\varepsilon, \eta} (-u) \, R_x  \rVert_{d+1, x}}{(1 + \frac{k}{2})^{d+1}}\,  \lVert A \rVert\, .
    \end{align*}
    Since $u \leq 1$ we can now use Lemma \ref{lem: cocycles} together with the fact that for all $\nu \in \N_0$ the supremum 
    $$\sup_{(\eps, \eta) \in [0,1]^2} \, \sup_{t \in I} \, \lVert \Tilde{S}_t^{\eps, \eta} \rVert_\nu $$
    is finite, to obtain for the above bound a constant $C_1 >0$ independent of $x$, $t$, $s$, $u$, $\eps$, $\eta$ and $A$, such that 
    \begin{align*}
        \hspace{2em}&\hspace{-2em} \sum_{x\in \Z^d \setminus X_c} \lVert [\, (1- \E_{B_{\frac{\dd(x,X)}{2}-\frac{c}{2}}(x)})\, \Tilde{\beta}_s^{\varepsilon, \eta} (-u) \, R_x ,\, \mU_{s, t}^{\eps, \eta} \, A \,] \rVert
        \\
        &\leq  C_1  \lvert X \rvert \, \lVert R \rVert_{d + 1} \, (1+ c)^{d-1}\, \lVert A \rVert 
        \, .
    \end{align*}
    The third part of our expression of interest can be bounded with the Lieb--Robinson bound of Assumption \hyperref[ass: LR bound]{(\ref*{ass: LR bound})}:
    \begin{align*}
        \hspace{2em}&\hspace{-2em}\sum_{x\in \Z^d \setminus X_c} \lVert [\, \E_{B_{\frac{\dd(x,X)}{2}-\frac{c}{2}}(x)} \, \Tilde{\beta}_s^{\varepsilon, \eta} (-u) \, R_x ,\, \mU_{s, t}^{\eps, \eta} \, A \,] \rVert
        \\
        &\leq \sum_{x\in \Z^d \setminus X_c} \frac{ \lvert X \rvert \, \lVert A \rVert \, \lVert R \rVert_0 \, f (\tfrac{1}{\eta} \, \lvert t -s \rvert)}{(1+ \max(0, (\frac{\dd(x,X)}{2} + \frac{c}{2})^{1/\kappa} - \frac{v}{\eta} \, \lvert t-s \rvert)^{\kappa(d+1)}}\, .
    \end{align*}
    At this point we choose $ c =  (2\, \frac{v }{\eta}\, \lvert t -s \rvert)^\kappa $ and use the concavity of $\R_+ \to \R_+, \, y \mapsto y^{1/\kappa}$ 
    \begin{align*}
        \hspace{2em}&\hspace{-2em}\sum_{x\in \Z^d \setminus X_c} \lVert [\, \E_{B_{\frac{\dd(x,X)}{2}-\frac{c}{2}}(x)} \, \Tilde{\beta}_s^{\varepsilon, \eta} (-u) \, R_x ,\, \mU_{s, t}^{\eps, \eta} \, A \,] \rVert
        \\
        &\leq \sum_{x\in \Z^d \setminus X_c} \frac{\lvert X \rvert \, \lVert A \rVert \, \lVert R \rVert_0 \, f (\tfrac{1}{\eta} \, \lvert t -s \rvert)}{(1+ \frac{1}{2}\dd(x,X)^{1/\kappa})^{\kappa(d+1)}}
        \\
        &\leq C_2\, \lvert X \rvert^2 \, \lVert R \rVert_0 \, \lVert A \rVert  \, f \left( \tfrac{1}{\eta} \, \lvert t -s \rvert \right)
        \, ,
    \end{align*}
    where $C_2>0$ is a constant independent of $t$, $s$, $\eps$, $\eta$ and $A$. Combining the three estimates and using that $\lvert X_c \rvert \leq \lvert X \rvert \, (1 + 2 \, c)^d$ yields the statement.
\end{proof}

\medskip
\begin{lemma} \label{Lemma: D_infty case}
    Let $\omega$ be a state on $\mA$ and $T : \mA_\infty \to \mA$ a linear operator. Assume that $\lim_{N \to \infty} \lVert T(\E_{B_N(x)} \,  A - A ) \rVert = 0$  for all $A \in \mA_\infty$, $x \in \Z^d$ and that there is $C > 0$ such that
    \begin{align*}
        |\omega(T \, A )| \leq C \, \lVert A \rVert \, |X|^2, \quad A \in \mA_X,
    \end{align*}
    for all $X \in P_0 (\Z^d)$. 
    Then there is another constant $\Tilde{C} > 0$ such that for all $A \in \mA_\infty$ and $x \in \Z^d$
    \begin{align*}
        |\omega(T \, A ) | \leq \Tilde{C} \, \lVert A \rVert_{2d+2, x} \, .
    \end{align*}
\end{lemma}

\begin{proof}
    We follow the proof of \cite[Lemma F.1]{Henheik_Teufel_2022} and set $A^{(0)} = \E_{B_1(x)} \, A$ and $A^{(j)}=(\E_{B_{j + 1} (x)} - \E_{B_j(x)}) \, A$ for $j \geq 1$. This defines a telescopic series and, therefore, $A = \sum_{j = 0}^\infty A^{(j)}$. Since $\lim_{N \to \infty} \w(T( \E_{B_N(x)} \, A)) = \w(T \, A)$, we observe 
    \begin{align*}
        \lvert \w(T \, A ) \rvert& \leq \sum_{j = 0}^\infty \lvert \w (T \, A^{(j)})\rvert
        \\
        &\leq \sum_{j = 1}^\infty C \, \lVert \left( \E_{B_{j+1} (x)} - \E_{B_j (x)} \right) \, A \rVert \, \lvert B_{j+1}(x) \rvert^2 + C \, \lVert \E_{B_1(x)} \, A \rVert \, \lvert B_1(x) \rvert^2
        \\
        & \leq C \left( \sum_{j = 1}^\infty \frac{1}{(1 + j)^{2d + 2}} \, \lvert B_{j+1}(x) \rvert^2 + \lvert B_1(x) \rvert^2 \right) \lVert A \rVert_{2d + 2, x} \, ,
    \end{align*}
    which gives us the desired bound.
\end{proof}

\section{The Inverse Liouvillian} \label{Appendix: Inverse Liouvillian}

In this section, we discuss the inverse Liouvillian as a mapping between spaces of interactions. It is primarily used for the quasi-local decomposition of operators and interactions into diagonal and off-diagonal parts with respect to a gapped ground state. In our proof  it appears at two points.
First, it is used to define the generator of the spectral flow automorphism \ref{lem: automorphic equivalence} and second, it provides us with a way of inverting the action of the Liouvillian inside the ground state expectation \ref{lem: liouvillian of inverse liouvillian}. We show that the well-known properties of the inverse Liouvillian hold in our setting and in particular also under the inclusion of Lipschitz potentials. See for example \cite{HastingsWen2005}, \cite{Bachmann2011}, \cite{NSY2019} for further information.

The following terminology will be useful for this section.
\medskip

\begin{definition}\label{def: differentiable path of gapped systems}
    Let $I \subseteq \R $ be an interval, $(H_t)_{t \in I} \in B_{\infty,I}^{(1)}$  a family of interactions,  $(\w_t)_{t \in I}$ a family of states such that \eqref{ass: gap of gs} and \eqref{ass: diff. of gs} are satisfied.
    Then we call the pair $((\w_t)_{t \in I}, (H_t)_{t \in I})$ a \emph{differentiable path of gapped systems}.
\end{definition}

\medskip

\begin{definition}\label{def: weight function}
    Let $((\w_t)_{t\in I}$, $(H_t)_{t\in I})$ be a differentiable path of gapped systems and let ${g_0}$ be the supremum of the set of all $c$, such that for all $t \in I$  the state $\w_t$ is a gapped ground state of $H_t$ with gap $c$. We define $g\coloneq \min \left( g_0, 1 \right)$ and the function $W \colon \R \to \R $ as the function from \cite[Lemma 2.6]{Bachmann2011}) with respect to $g$. We note the properties of $W$ that are relevant here:

    \begin{itemize}
        \item[(i)] $W$ is odd.
        \item[(ii)] The Fourier transform $\widehat W$ of $W$ satisfies $\widehat{W}(k) = \frac{- \i}{\sqrt{2  \pi}\, k}$ for  $k \in \R \setminus [-g,g]$.
        \item[(iii)] $\sup_{s\in\R} \lvert s \rvert^n \, W(s) < \infty$ for all $n\in \N_0$.
    \end{itemize}
\end{definition}

\medskip

\begin{definition}
    Let $((\w_t)_{t \in I}, \, (H_t)_{t \in I} )$ be a differentiable path of gapped systems and $\Psi = \Phi + V$, where $\Phi$ is a $B_\infty$-interaction and $V$ a Lipschitz potential. We define the quasi-local operators
    \begin{align*}
        \mI_t (\Psi)_{x,*} \coloneq - \i  \int_{\R}\dd s \, W(s) \int_0^s \dd u \, \e^{\i u \mL_{H_t}} \,  \mL_{\Psi}  \, H_{t, x} \, ,
    \end{align*}
    and
    \begin{align*}
        (\Psi\OD[t])_{x,*} \coloneq \i \int_{\R}\dd s \, W(s)\, \e^{\i s \mL_{H_t}} \, \mL_{\Psi} \, H_{t, x} \, ,
    \end{align*}
    for each $x\in \Z^d$ and the interactions $\mI(\Psi)$, $\Psi\OD$ by
    \begin{align*}
        \mI_t(\Psi)(B_k(x)) &= (\E_{B_k(x)} - \E_{B_{k-1}(x)})\, \mI_t(\Psi)_{x,*}\\
        \mI_t(\Psi)(B_0(x)) &= \E_{B_0(x)}\, \mI_t(\Psi)_{x,*}
    \end{align*}
    and
    \begin{align*}
        \Psi\OD[t](B_k(x)) &= (\E_{B_k(x)} - \E_{B_{k-1}(x)})\, (\Psi\OD[t])_{x,*}\\
        \Psi\OD[t](B_0(x)) &= \E_{B_0(x)}\, (\Psi\OD[t])_{x,*}
    \end{align*}
    for all $(x,k) \in \Z^d \times \N$ and as $\mI_t(\Psi)(M) = \Psi\OD[t] (M) = 0$ for all $M \in P_0(\Z^d)$ that are not equal to $B_k(x)$ for some $(x,k) \in \Z^d \times \N_0$. We call the interactions $\mI_t(\Psi)$ \emph{inverse Liouvillian of the interaction $\Psi$} and the interaction $\Psi\OD[t]$ \emph{off-diagonal part of the interaction $\Psi$}
\end{definition}

\medskip
\begin{remark}
    Note that the inverse Liouvillian and the off-diagonal part of an interaction is defined such that for all $x \in \Z^d$ it holds that $\mI_t(\Psi)_x = \mI_t(\Psi)_{x, *}$ and $(\Psi\OD[t])_x = (\Psi\OD[t])_{x, *}$.
\end{remark}

\medskip
\begin{lemma}[{\cite[Lemma C.4]{Becker_Teufel_Wesle_2025}}] \label{Lemma: Lieb-Robinson bound for D_infty}
    Let $((\w_s)_{s\in I}, (H_s)_{s\in I})$ be a differentiable path of gapped systems and $A\in \mA_\infty$. For each $\nu\in \N_0$ there exists an increasing and at most polynomially growing function  function $ b_\nu \colon [0,\infty) \to [0,\infty)$, such that for all $t\in \R$ and $z \in \Z^d$, it holds that
    \begin{align*}
        \sup_{s\in I} \left\lVert \e^{\i t \mL_{H_s}} \, A \right\rVert_{\nu,z} 
        \leq b_\nu(\lvert t \rvert) \, \lVert A \rVert_{\nu,z} \, .
    \end{align*}
\end{lemma}

\medskip
\begin{lemma} \label{lem: liouvillian of inverse liouvillian}
    Let $((\w_t)_{t \in I}, \, (H_t)_{t \in I})$ be a differentiable path of gapped systems, and $\Psi = \Phi + V$, where $\Phi \in B_\infty$ and $V \in \mV$. For all $A \in \mA_\infty$ it holds that
    \begin{align*}
        \w_t(\mL_\Psi \, A) = \w_t(\mL_{\Psi\OD[t]} \, A) 
    \end{align*}
    and
    \begin{align*}
        \mL_{- \i [H_t, \mI_t (\Psi)]} \, A = \mL_{\Psi\OD[t]} \, A \, .
    \end{align*}
\end{lemma}
\begin{proof}
    For the first claim, we use Lemma \ref{lem: sum representation of generator} and absolute convergence due to Lemmas \ref{Lemma: Lieb-Robinson bound for D_infty} and \ref{lem:commutator bound} to exchange integration and summation. We find
    \begin{align*}
        \w_t (\mL_{\Psi\OD[t]} \, A) &= \sum_{x \in \Z^d} \w_t( [\i \int_{\R}\dd s \, W(s)\, \e^{\i s \mL_{H_t}} \, \mL_{\Psi} \, H_{t, x}, \, A ] )
        \\
        &= - \sum_{y \in \Z^d} \w_t([ \i \int_{\R}\dd s \, W(s)\, \e^{\i s \mL_{H_t}} \, \mL_{H_t} \, \Psi_{y}, \, A] )
        \\
        &= - \sum_{y \in \Z^d} \w_t( [ \i \mL_{H_t} \, \mI_{H_t} \, \Psi_{y}, \, A ] ) \, ,
    \end{align*}
    where $\mI_{H_t}$ is defined as in \cite{henheik2022adiabatic} by 
    \begin{align*}
        \mI_{H_t} \, A \coloneq \int_\R \d s \, W(s) \, \e^{\i s \mL_{H_t}} \, A \, ,
    \end{align*}
    for $A \in \mA$. Note that \cite[Proposition 3.3]{henheik2022adiabatic} (which in turn relies on \cite[Proposition 6.9]{NSY2019}) also applies to our setting, where $H_t$ is only super-polynomially decaying, since this property was not used in the proof. Using this proposition we obtain
    \begin{align*}
        \w_t(\mL_{\Psi\OD[t]} \, A) &= \sum_{y \in \Z^d} \w_t([\Psi_y, \, A])
        \\
    &= \w_t(\mL_\Psi \, A) \, ,
    \end{align*}
    showing the first statement. For the second statement, we use Lemma \ref{lem: sum liouvillian interaction} and obtain
    \begin{align*}
        \mL_{- \i \, [H_t, \, \mI_t(\Psi)]} \, A &= \i \, \sum_{x \in \Z^d} \int_\R \d s \, W(s) \int_0^s \d u \, [ \i \mL_{H_t} \, \e^{\i u \mL_{H_t}} \, \mL_\Psi \, H_{t, x}, \, A]
        \\
        & = \i \, \sum_{x \in \Z^d} \int_\R \d s \, W(s) \, [ \e^{\i s \mL_{H_t}} \, \mL_\Psi \, H_{t, x}, \, A]
        \\
        &= \sum_{x \in \Z^d} [(\Psi\OD[t])_x , \, A]
        \\
        &= \mL_{\Psi\OD[t]} \, A \, ,
    \end{align*}
    where absolute convergence to excahnge integration and summation follows from the same lemmas.
\end{proof}

\medskip
\begin{lemma} \label{lem: inverse liouvillian of b_infty}
    Given a differentiable path of gapped systems $((\w_t)_{t \in I}, (H_t)_{t \in I})$. If one has interactions $(\Phi_t)_{t \in I} \in B_{\infty, I, L}^{(k)}$ and $(V_t)_{t \in I} \in \mV_{I, L}^{(k)}$ for some $k \in \N_0$ and $L \subseteq \Z^d$, then $(\mI_t(\Psi_t))_{t\in I} \in B^{(k)}_{\infty, I}$ and $(\Psi_t\OD[t])_{t \in I} \in B_{\infty, I}^{(k)}$, where $(\Psi_t)_{t \in I} = (\Phi_t + V_t)_{t \in I}$.
    
    Moreover, for any $\nu, n \in \Z^d$ and $m \leq k$, it holds that
    \begin{align*}
        \sup_{t \in I} \sup_{x \in \Z^d} \, \left\| \left( \frac{\d^m}{\d t^m} \, \mI_t (\Psi_t) \right)_{x} \right\|_{\nu, x} \, (1 + \d (x, L))^n < \infty \, .
    \end{align*}
\end{lemma}
\begin{proof}
We separate the cases for $\Phi_t$ and $V_t$. Showing that $(\mI_t(\Phi_t))_{t \in I} \in B_{\infty, I}^{(0)}$ follows from \cite[Lemma B.6]{Becker_Teufel_Wesle_2025}. Estimates for higher derivatives can be performed in the same way.

We now prove the statement for a family of Lipschitz potentials $(V_s)_{s \in I}$ and $k = 0$. We first note that for all $\nu \in \N_0$ we have $\sup_{x \in \Z^d }\lVert H_{s,x} \rVert_{\nu,x} \leq 3 \, \lVert H_s \rVert_\nu$. Hence, we find by Lemma \ref{Lemma: Lieb-Robinson bound for D_infty} and Lemma \ref{lem: sum representation of generator} that $\sup_{s \in I} \sup_{ x \in \Z^d} \lVert \mI_s (V_s)_x \rVert_{\nu , x} < \infty$. We calculate
    \begin{align*}
        \lVert \mI_s( V_s ) \rVert_\nu &= \sup_{x \in \Z^d} \, \sum_{ \substack{M \in P_0(\Z^d) \\ x \in M }} (1 + \diam (M) )^\nu \, \lVert \mI_s(V_s)(M) \rVert\\
        &\leq \sup_{x \in \Z^d} \, (  \sum_{ k = 1 }^\infty \, \sum_{ y \in B_k(x)} (1 + 2k )^\nu \, \lVert (\E_{ B_k ( x ) } - \E_{B_{k-1}( x ) } ) \, \mI_s ( V_s )_y \rVert + \lVert \E_{B_0(x)} \, \mI_s ( V_s )_x \rVert) \\
        &\leq \sum_{ k = 1 }^\infty \frac{(1 + 2 k )^d \, (1 + 2k )^\nu}{(1 + (k - 1) )^{\nu + d + 2}} \,  \sup_{x \in \Z^d} \, \lVert  \mI_s ( V_s )_x \rVert_{\nu + d + 2, x} + \sup_{x \in \Z^d} \, \lVert \mI_s ( V_s )_x \rVert \, ,    
    \end{align*}
    which gives $\sup_{s\in I}\lVert \mI_s(V_s) \rVert_\nu < \infty$ for all $\nu \in \N_0$. 
    
    To see continuity of $\mI_s (V_s) (M)$ for all $M \in P_0(\Z^d)$, we notice that
    \begin{align*}
        \lVert \mI_{s+h} (\Phi_{s+h}) (M) - \mI_s (\Phi_s) (M) \rVert \leq 2 \, \lVert \mI_{s+h} (\Phi_{s+h})_x - \mI_s (\Phi_s)_x \rVert \, ,
    \end{align*}
    for some $x \in \Z^d$. Now compute   
    \begin{align*}
       \lVert \mI_{s+h} (V_{s+h})_x - \mI_{s} (V_{s})_x \rVert \leq \int_\R \dd t \, W(t) \int_0^t \dd u \, \lVert \e^{\i u \mL_{H_{s+h}}} \mL_{ V_{s+h} } \, H_{s+h,x} - \e^{\i u \mL_{H_s}} \mL_{ V_{s} } \, H_{s,x} \rVert \, .   
    \end{align*}
    We split the integrand as follows
    \begin{align*}
        \lVert \e^{\i u \mL_{H_{s+h}}} \mL_{ V_{s+h} } \, H_{s+h,x} - \e^{\i u \mL_{H_s}} \mL_{ V_{s} } \, H_{s,x} \rVert 
        & \leq \lVert \mL_{V_{s+h}}\, (H_{s+h,x} - H_{s,x} ) \rVert \\
        &\quad + \lVert  (\mL_{V_{s+h}} - \mL_{V_s}) \, H_{s,x} \rVert \\
        & \quad + \lVert ( \e^{\i u \mL_{H_{s+h}}} - \e^{\i u \mL_{H_s}}) \, \mL_{V_s} \, H_{s,x} \rVert \, .
    \end{align*}
    The first term vanishes in the limit $h \to 0$ due to Lemmas \ref{lem: sum representation of generator} and \ref{lem: derivative of zero chain}. The second term vanishes in the limit $h \to 0$ by Lemma \ref{lem: derivative of Liouvillian}. For the last term we observe that by Duhamel's formula
    \begin{align*}
        \lVert (\e^{ \i u \mL_{H_{s + h}}} - \e^{\i u \mL_{H_s}}) \, \mL_{V_s} \, H_{s,x} \rVert &\leq \int_0^h \d t \, \lVert \partial_t \, \e^{\i u \mL_{H_{s + t}}} \, \mL_{V_s} \, H_{s,x} \rVert
        \\
        & \leq \int_0^h \d t \,  \int_0^u \d v \, \lVert \e^{\i v \mL_{H_{s + t}}} \, \mL_{\Dot{H}_{s+t}} \, \e^{\i (u - v) \mL_{H_{s + t}}} \, \mL_{V_s} \, H_{s,x} \rVert \, .
    \end{align*}    
    This term also vanishes in the limit $h \to 0$ by Lemma \ref{lem: sum representation of generator} and Lemma \ref{Lemma: Lieb-Robinson bound for D_infty}. The claim then follows by dominated convergence. The cases for $k > 0$ are done in the same way.

    The statement for the off-diagonal part follows by the same argument.

    To see the localization statement for $k = 0$, we calculate using Lemma \ref{Lemma: Lieb-Robinson bound for D_infty}
    \begin{align*}
        \lVert (\mI_t (\Psi_t))_x \rVert_{\nu, x} \, (1 + \d (x, L))^n &\leq \int_\R \d s \, \lvert W(s) \rvert \int_0^s \d u \, \lVert \e^{\i u \mL_{H_t}} \, \mL_{\Psi_t} \, H_{t, x} \rVert_{\nu, x} \, (1 + \d (x, L))^n
        \\
        & \leq C \, \lVert \mL_{\Psi_t} \, H_{t, x} \rVert_{\nu, x} \, (1 + \d (x, L))^n \, ,
    \end{align*}
    where $C$ is some constant. This expression is uniformly bounded in $x \in \Z^d$ and $t \in I$ by assumption. 

    For $k = 1$ we realize that using Duhamel's formula, Lemma \ref{lem: derivative of zero chain}, and Lemma \ref{lem: derivative of Liouvillian} we get
    \begin{align*}
        \left\| \left(\frac{\d}{\d t} \, \mI_t (\Psi_t) \right)_x \right\|_{\nu, x} \, (1 + \d (x, L))^n &\leq \int_\R \d s \, \lvert W(s) \rvert \int_0^s \d u \, \Big( \lVert \e^{\i u \mL_{H_t}} \, \mL_{\Psi_t} \, \Dot{H}_{t, x} \rVert_{\nu, x} 
        \\
        & \qquad + \lVert \e^{\i u \mL_{H_t}} \, \mL_{\Dot{\Psi}_t} \, H_{t, x} \rVert_{\nu, x}
        \\
        & \qquad + \int_0^u \d v \, \lVert \e^{\i v \mL_{H_t}} \, \mL_{\Dot{H}_t} \, \e^{\i  (u - v) \mL_{H_t}} \mL_{\Psi_t} \, H_{t, x} \rVert_{\nu, x} \Big) \, (1 + \d (x, L))^n \, .
    \end{align*}
    These three terms can be bounded in the same way as above. The statement for higher derivatives follows in the same way.
\end{proof}

\medskip
\begin{lemma} \label{lem: automorphic equivalence}
    Given a differentiable path of gapped systems $((\w_t)_{t \in I}, \, (H_t)_{t \in I})$ and $A \in \mA_\infty$ be arbitrary. Then
    \begin{align*}
        \Dot{\w}_t (A) = -\i \w_t(\mL_{\mI_t( \Dot{H}_t) } \, A ) \, .
    \end{align*}
\end{lemma}
\begin{proof}
    We notice that in our setting all assumptions of \cite[Theorem 3.2]{Becker_Teufel_Wesle_2025} are satisfied. This theorem gives us that ground states at different times are connected by an automorphism generated by the family of interactions $ (- \mI_t (\Dot{H}_t))_{t \in I}$. Taking the derivative yields the lemma.
\end{proof}

\section{Technical Lemmas}\label{sec: techincal lemmas}

\smoothcommutator*
\begin{proof}[Proof of Lemma \ref{lem: comm is smooth}]\label{proof:smoothcommutator}
    We divide the proof into two parts. We first show that the commutator lies in $B_{\infty, I}^{(k)}$, and then that it is in $B_{\infty, I, L}^{(k)}$. 
    
    First let $k = 0$, that is $(\Phi_t)_{t \in I} \, , \, (\Psi_t)_{t \in I} \in B_{\infty, I}^{(0)}$ and $(V_t)_{t \in I} \in \mV_{I}^{(0)}$. By Lemma \ref{lem: Commutator of interaction} we know that $\i [\Phi_t + V_t, \, \Psi_t]$ is a $B_\infty$ interaction for all $t \in I$. Furthermore, the lemma gives us the bound
    \begin{align*}
        \sup_{t \in I} \, \lVert \i \, [\Phi_t + V_t, \, \Psi_t] \rVert_{\nu} \leq 2^{d + 2} \, \sup_{t \in I} \lVert \Psi_t \rVert_{\nu + d} \, \sup_{t \in I} \lVert \Phi_t \rVert_{\nu + d} + 3 \sup_{t \in I} C_{v_t} \, \sup_{t \in I} \lVert \Psi_t\rVert_{\nu + d + 2} \, .
    \end{align*}
    Now let $M \in P_0(\Z^d)$ be arbitrary. To see the continuity of the map $t \mapsto \i \, [\Phi_t + V_t, \, \Psi_t](M)$ we observe that
    \begin{align*}
        \i \, [\Phi_t + V_t, \, \Psi_t](M) = \i  \sum_{\substack{M_1, M_2 \subseteq M \\ M_1 \cup M_2 = M}} [\Phi_t (M_1) + V_t (M_1), \, \Psi_t(M_2)] \, 
    \end{align*}
    is a finite sum. We can therefore pull any limit through it. Since the commutator is given by sums and products of continuous maps $I \to \mA_0^N$, the map $t \mapsto \i \, [\Phi_t + V_t, \, \Psi_t](M)$ is also continuous. The statement for $k > 0$ can be shown in the same way.

    Now, we prove the second claim. To this end, observe that using Lemmas \ref{lem: sum liouvillian interaction} and \ref{lem:commutator bound} for any $n, \nu \in \N$ and $A \in \mA_\infty$ it holds that
    \begin{align*}
        \hspace{6ex} & \hspace{-6ex} \lVert \mL_{\i [\Phi_t + V_t, \Psi_t]} \, A \rVert_{\nu, x} \, (1 + \d (x, L))^n 
        \\
        & \leq \sum_{y \in \Z^d} \, \lVert [ \mL_{\Phi_t + V_t} \, (\Psi_t)_y, \, A] \rVert_{\nu, x} \, (1 + \d (x, L))^n 
        \\
        & \leq \sum_{y \in \Z^d} 4^{\nu + n + d + 4} \, \frac{ \lVert \mL_{\Phi_t + V_t} \, (\Psi_t)_y \rVert_{\nu + n + d + 1, y} \, \lVert A \rVert_{\nu + n + d + 1, x}}{(1 + \lVert x - y \rVert)^{n + d + 1}} (1 + \d (x, L))^n
        \\
        & \leq \sum_{y \in \Z^d} 4^{\nu + n + d + 4} \, \frac{ \lVert \mL_{\Phi_t + V_t} \, (\Psi_t)_y \rVert_{\nu + n + d + 1, y} \, \lVert A \rVert_{\nu + n + d + 1, x}}{(1 + \lVert x - y \rVert)^{n + d + 1}} (1 + \d (y, L))^n \, (1 + \d (y, x))^n
        \\
        & \leq \sup_{z \in \Z^d} \, (\lVert \mL_{\Phi_t + V_t} \, (\Psi_t)_z \rVert_{\nu + n + d + 1, z} \, (1 + \d (z, L))^n ) \sum_{y \in \Z^d} 4^{\nu + n + d + 4} \, \frac{ \lVert A \rVert_{\nu + n + d + 1, x} }{(1 + \lVert x - y \rVert)^{d + 1}}.
    \end{align*}
    This shows that $(\i \, [\Phi_t + V_t, \Psi_t])_{t \in I} \in B_{\infty, I, L}^{(0)}$. This proof carries over to the case of higher derivatives, since the derivative of a commutator of $L$-localized interactions is also a commutator of $L$-localized interactions.
\end{proof}

\sumrepresentation*
\begin{proof}[Proof of Lemma \ref{lem: sum representation of generator}]\label{proof:sumrepresentation}
    The statement with only the $B_\infty$-interaction $\Phi$ can be found in \cite[Lemma 2.7]{Becker_Teufel_Wesle_2025} where the bound is slightly worse. To see the improved bound we simply estimate with Lemma \ref{lem:commutator bound}. Let $\nu\in\N_0$ and $x \in \Z^d$. We find
    \begin{align*}
        \sum_{x\in \Z^d} \lVert [\, \Phi_x ,\, A \,] \rVert_{\nu,x} 
        &\leq 4^{\nu+d+4} \sum_{z\in \Z^d} \frac{\lVert \Phi_z \rVert_{\nu+d+1,z} \, \lVert A \rVert_{\nu+d+1,x} \, }{(1+\lVert z-x \rVert)^{d+1}}
        \\
        & \leq 4^{\nu+d+4} \sum_{z\in \Z^d} \frac{3\, \lVert \Phi \rVert_{\nu+d+1} \, \lVert A \rVert_{\nu+d+1,x} \, }{(1+\lVert z-x \rVert)^{d+1}}\, .
    \end{align*}
    To see the statement for Lipschitz potentials let $v: \Z^d \to \R$ be such that $V(\{x\}) = v(x)\, n_x$. We first consider the shifted sum 
    \begin{align*}
        \sum_{z\in \Z^d}  [\,(v(z)-v(x))\, n_z,\, A\, ]\, .
    \end{align*}
    Lemma \ref{lem:commutator bound} gives us 
    \begin{align*}
        \sum_{z\in \Z^d} \lVert [\,(v(z)-v(x))\, n_z,\, A\, ] \rVert_{\nu,x}
        & \leq \sum_{z\in \Z^d} 4^{\nu + d +5} \, \frac{\lvert v(z) - v(x)\rvert \, \lVert n_z \rVert_{\nu+d+2,z} \, \lVert A \rVert_{\nu+d+2,x} }{(1 + \lVert z-x \rVert)^{d+2}}
        \\
        & \leq 4^{\nu + d + 5} \, C_v \, \lVert n_0 \rVert  \sum_{z\in \Z^d}  \frac{\lVert z - x\rVert   }{(1 + \lVert z-x \rVert)^{d+2}} \,  \lVert A \rVert_{\nu+d+2,x}
        \, ,
    \end{align*}
    showing absolute convergence. Since the number operator $N$ is in $B_\infty$ we can use the first claim to see that
    \begin{align*}
        \sum_{z\in \Z^d}  [\,v(z)\, n_z,\, A\, ] 
        & = \sum_{z\in \Z^d}  [\,(v(z)-v(x))\, n_z,\, A\, ] + v(x)\sum_{z\in \Z^d}  [\,n_z,\, A\, ]
    \end{align*}
    is the sum of two absolutely convergent sums and thus absolutely convergent. Thus we have
    \begin{align*}
        \lVert \sum_{z\in \Z^d}  [\,v(z)\, n_z,\, A\, ] \rVert_{\nu,x} 
        &\leq 4^{\nu + d + 5} \, C_v \, \lVert n_0 \rVert  \sum_{z \in \Z^d}  \frac{\lVert z - x\rVert   }{(1 + \lVert z-x \rVert)^{d + 2}} \,  \lVert A \rVert_{\nu+d+2,x} 
        + \lvert v(x) \rvert \, \lVert \mL_N\, A \rVert_{\nu,x}
        \\
        & \leq 4^{\nu + d + 5} \, C_v \, \lVert n_0 \rVert  \sum_{z \in \Z^d}  \frac{\lVert z - x\rVert   }{(1 + \lVert z-x \rVert)^{d + 2}} \,  \lVert A \rVert_{\nu+d+2,x}
        \\
        & \quad + 4^{\nu+d+4}\, \lvert v(x) \rvert \, \lVert n_0 \rVert \sum_{z\in\Z^d} \frac{1}{(1+\lVert z -x\rVert)^{d+1}} \lVert A \rVert_{\nu+d+1,x}
        \, .
    \end{align*}
    This also establishes that the tuple 
    \begin{align*}
        (\E_{B_k(0)}\, A, \, \mL^\circ_{V} \, \E_{B_k(0)}\, A)
    \end{align*}
    converges to 
    \begin{align*}
        (A, \sum_{z\in \Z^d} [\, v(z) \, n_z,\, A\, ] ) \, ,
    \end{align*}
    as $k \to \infty$. Since $\mL_V$ is the closure of $\mL_V^\circ$, this means that $A\in D(\mL_V)$ and 
    \begin{align*}
        \mL_V\,A = \sum_{z\in \Z^d} [\, v(z) \, n_z ,\, A \,]
        \,.
    \end{align*}
    In particular the bound 
    \begin{align*}
        \lVert \mL_V\, A \rVert_{\nu,x}
        & \leq 4^{\nu + d + 5} \, C_v \, \lVert n_0 \rVert  \sum_{z \in \Z^d}  \frac{\lVert z - x\rVert   }{(1 + \lVert z-x \rVert)^{d + 2}} \,  \lVert A \rVert_{\nu+d+2,x}
        \\
        & \quad + 4^{\nu+d+4}\, \lvert v(x) \rvert \, \lVert n_0 \rVert \sum_{z\in\Z^d} \frac{1}{(1+\lVert z -x\rVert)^{d+1}} \lVert A \rVert_{\nu+d+1,x}
    \end{align*}
    holds and if $A$ is gauge-invariant, since $\mL_N \, A = 0$, we have the stronger bound
    \begin{align*}
        \lVert \mL_V\, A \rVert_{\nu,x} 
        \leq 4^{ \nu + d+5} \, \lVert n_0 \rVert  \sum_{z \in \Z^d}  \frac{\lVert z - x\rVert   }{(1 + \lVert z-x \rVert)^{d+2}} \, C_v \, \lVert A \rVert_{\nu+d+2,x}\, .
    \end{align*}
    Combining the bounds gives the lemma.
\end{proof}

\medskip
\begin{lemma} \label{lem: Commutator of interaction}
    Let $\Psi$, $\Phi \in B_\infty$ and $V \in \mV$ with Lipschitz constant $C_v$. For all $\nu\in \N_0$ it holds that
    \begin{align*}
        \lVert \i[\,\Psi + V ,\, \Phi\,] \rVert_{\nu} \leq 2^{d + 2} \, \lVert \Psi \rVert_{\nu + d} \, \lVert \Phi \rVert_{\nu +d} + 3 \, C_v \, \lVert \Phi \rVert_{\nu + d + 2} \, .
    \end{align*}
    In particular, $\i [\Psi + V, \, \Phi]$ is a $B_\infty$ interaction.
\end{lemma}
\begin{proof}
    The case for $\i \, [\Psi, \, \Phi]$ was done in \cite[Proposition 2.13]{wmmmt2024}. Note that for the proof of the bound, periodicity was not required. Therefore we have
    \begin{align*}
    \lVert \i \, [\Psi, \, \Phi] \rVert_\nu \leq 2^{d + 2} \, \lVert \Psi \rVert_{\nu + d} \, \lVert \Phi \rVert_{\nu +d} \, .
    \end{align*}
    Since Lipschitz potentials are only supported on single element sets, we find
    \begin{align*}
        \i \, [V, \, \Phi](M) &= \sum_{x \in M} \i \, [v(x) \, n_x, \, \Phi(M)]
        \\
        & = \sum_{x \in \Z^d} \i \, [v(x) \, n_x, \, \Phi(M)]
        \\
        & = \i \mL_{ V} \, \Phi(M) \, .
    \end{align*}
    By Lemma \ref{lem: sum representation of generator} we find
    \begin{align*}
        \lVert \i \, [V, \, \Phi] \rVert_\nu &\leq \sup_{x \in \Z^d} \sum_{\substack{M \in P_0(\Z^d) \\ x \in M}} (1 + \mathrm{diam}(M) )^\nu \, \lVert \mL_V \, \Phi(M) \rVert \\
        &\leq \sup_{x \in \Z^d} \sum_{\substack{M \in P_0(\Z^d) \\ x \in M}} (1 + \mathrm{diam}(M) )^\nu \, C_v \, \lVert \Phi(M) \rVert_{d + 2, x}\\
        &\leq \sup_{x \in \Z^d} \sum_{\substack{M \in P_0(\Z^d) \\ x \in M}} 3 \, (1 + \mathrm{diam}(M) )^{\nu + d + 2} \, C_v \, \lVert \Phi(M) \rVert \, . \qedhere
    \end{align*}
\end{proof}

\medskip
\begin{lemma} \label{lem: sum liouvillian interaction}
    Let $\Psi, \Phi \in B_\infty$, $V \in \mV$, and $A \in \mA_\infty$, then 
    \begin{align*}
        \mL_{\i[\Psi + V,\Phi]}\, A = \sum_{x\in \Z^d} [\,\i\mL_{\Psi + V} \, \Phi_x,\, A\,] \, .
    \end{align*}
\end{lemma}
\begin{proof}
    We first prove the case for the $B_\infty$ interaction. Using Lemma \ref{lem: Commutator of interaction} and Lemma \ref{lem: sum representation of generator} we have
    \begin{align*}
        \mL_{\i [\Psi , \Phi]} \, A = \sum_{x \in \Z^d} \sum_{M \in R_x(\Z^d)} \sum_{\substack{M_1, M_2 \subseteq M \\ M_1 \cup M_2 = M}} [\i \, [\Psi(M_1), \, \Phi(M_2)], \, A] \, .
    \end{align*}
    Using Lemma \ref{lem:commutator bound}, we find 
    \begin{align*}
        \hspace{4ex} & \hspace{-4ex} \sum_{x \in \Z^d} \sum_{M \in R_x(\Z^d)} \sum_{\substack{M_1, M_2 \subseteq M \\ M_1 \cup M_2 = M}} \lVert [\i \, [\Psi(M_1), \, \Phi(M_2)], \, A] \rVert
        \\
        & \leq \sum_{x \in \Z^d} \sum_{M \in R_x(\Z^d)} \sum_{\substack{M_1, M_2 \subseteq M \\ M_1 \cup M_2 = M}} 4^{d + 4} \, \frac{\lVert \i \, [\Psi(M_1), \, \Phi(M_2)] \rVert_{d + 1, x} \, \lVert A \rVert_{d + 1, y}}{(1 + \lVert x - y \rVert)^{d + 1}} \, .
    \end{align*}
    To see that this term is bounded, notice that
    \begin{align*}
        \hspace{4ex} & \hspace{-4ex} \sum_{M \in R_x(\Z^d)} \sum_{\substack{M_1, M_2 \subseteq M \\ M_1 \cup M_2 = M}} \lVert \i \, [\Psi(M_1), \, \Phi(M_2)] \rVert_{d + 1, x} 
        \\
        & \leq \sum_{\substack{M \in P_0(\Z^d) \\ x \in M}} \sum_{\substack{M_1, M_2 \subseteq M \\ M_1 \cup M_2 = M}} 3 \, (1 + \mathrm{diam}(M))^{d + 1} \lVert \i \, [\Psi(M_1), \, \Phi(M_2)] \rVert \, .
    \end{align*}
    This term can be bounded in the same way as in the proof of \cite[Proposition 2.13]{wmmmt2024}. Hence, the sum is absolutely convergent and we can reorder the summands. We find
    \begin{align*}
        \mL_{\i [\Psi, \Phi]} \, A &= \sum_{x \in \Z^d} \sum_{M_1 \in P_0(\Z^d)} \sum_{M_2 \in R_x(\Z^d)} [\i \, [\Psi(M_1), \, \Phi(M_2)], \, A]
        \\
        & = \sum_{x \in \Z^d} \sum_{M_1 \in P_0(\Z^d)} [\i \, [\Psi(M_1), \, \Phi_x], \, A]
        \\
        & = \sum_{x \in \Z^d} [\i \, \mL_{\Psi} \, \Phi_x, \, A] \, ,
    \end{align*}
    where we used Lemma \ref{lem: sum representation of generator} in the last equality.

    For the Lipschitz potential using Lemma \ref{lem: Commutator of interaction} and Lemma \ref{lem: sum representation of generator} we find 
    \begin{align*}
        \mL_{\i [V, \Psi]} \, A &= \sum_{x \in \Z^d} [\i \sum_{M \in R_x} \, \sum_{y \in M}[v(y) \, n_y, \, \Psi(M)], \, A] \\
        &= \sum_{x \in \Z^d} [\i \sum_{M \in R_x} \, \sum_{y \in \Z^d}[v(y) \, n_y, \, \Psi(M)], \, A] \, .
    \end{align*}
    To see the absolute convergence of the sums we use Lemma \ref{lem:commutator bound} and calculate
    \begin{align*}
        \hspace{4ex} & \hspace{-4ex} \sum_{x \in \Z^d} \sum_{M \in R_x} \, \sum_{y \in \Z^d} \lVert [ \i \, [(v(y)- v(0) ) \, n_y, \, \Psi(M)], \, A] \rVert \\
        & \leq \sum_{x \in \Z^d} \sum_{M \in R_x} \, \sum_{y \in \Z^d} 4^{ d + 5} \frac{\lVert [(v(y) - v(0)) \, n_y, \, \Psi(M)] \rVert_{d + 2, y} \, \lVert A \rVert_{d + 2, 0}}{(1 + \lVert y \rVert )^{d + 2}} \\
        & \leq \sum_{x \in \Z^d} \sum_{M \in R_x} \, \sum_{y \in \Z^d} 4^{3d + 11} \, C_v \, \frac{\lVert y \rVert \, \lVert n_y \rVert_{2d + 3, y} \, \lVert \Psi(M) \rVert_{2d + 3, x} \, \lVert A \rVert_{d + 2, 0}}{(1 + \lVert y \rVert )^{d + 2} \, (1 + \lVert x - y \rVert)^{d + 1}}\\
        & \leq \sum_{x \in \Z^d} \sum_{M \in R_x} \, \sum_{y \in \Z^d} 4^{3d + 11} \, C_v \, 3 (1 + \diam (M))^{2d + 3} \, \frac{\lVert y \rVert \, \lVert n_y \rVert_{2 d + 3, y} \, \lVert \Psi(M) \rVert \, \lVert A \rVert_{d + 2, 0}}{(1 + \lVert y \rVert )^{d + 2} \, (1 + \lVert x - y \rVert)^{d + 1}} < \infty \, .       
    \end{align*}
    Hence, we can reorder the summands and find 
    \begin{align*}
        \mL_{\i [V, \Psi]} \, A  &= \sum_{x \in \Z^d} [\i \, \sum_{y \in \Z^d} [v(y) \, n_y, \, \Psi_x], \, A ] \\
        &= \sum_{x \in \Z^d} [\i \, \mL_V \, \Psi_x, \, A] \, . \qedhere
    \end{align*}
\end{proof}

\medskip
\begin{lemma}[{\cite[Lemma C.1]{Becker_Teufel_Wesle_2025}}] \label{lem: derivative of zero chain}
    Let $I \subseteq \R$ be some interval and $(\Phi_s)_{s \in I} \in B_{\infty, I}^{(k)}$. The map $s \mapsto \Phi_{s, x}$ is $k$ times continuously differentiable with respect to $\lVert \cdot \rVert _{\nu, x}$ for all $\nu \in \N_0$ and $x \in \Z^d$. For $0\leq m \leq k$ the derivative is given by $\frac{\d^m}{\d s^m} \Phi_{s,x}  =(\frac{\d^m}{\d s^m} \Phi_s)_x$, where $\frac{\d^m}{\d s^m}\Phi_s$ denotes the interaction obtained by term-wise differentiation.
\end{lemma}

\medskip

\begin{lemma}\label{lem: derivative of Liouvillian}
Let $I \subseteq \R$ be some interval, $(\Phi_s)_{s \in I} \in B_{\infty, I}^{(k)}$, $(V_s)_{s \in I} \in \mV_I^{(k)}$ and $A \in \mA_\infty^N$. The map $s \mapsto \mL_{\Phi_s + V_s} \, A$ then is $k$ times continuously differentiable with respect to $\lVert \, \cdot \, \rVert_{\nu, x}$ for any $x \in \Z^d$ and $\nu \in \N_0$. For all $0 \leq m \leq k$ the derivative is given by $\frac{\d^m}{\d s^m} \, (\mL_{\Phi_s + V_s} \, A) = \mL_{\frac{\d^m}{\d s^m} \Phi_s + \frac{\d^m}{\d s^m} V_s} \, A$, where $\frac{\d^m}{\d s^m}\Phi_s$ and $\frac{\d^m}{\d s^m} V_s$ denote the interactions obtained by term-wise differentiation.
\end{lemma}
\begin{proof}
    Since the Liouvillian is linear in its arguments, we can split the proof into a part involving the $B_{\infty, I}^{(k)}$ interaction and one with the $\mV_I^{(k)}$ interaction. The case for the $B_\infty$ interaction was done in \cite[Lemma C.2]{Becker_Teufel_Wesle_2025}.
    
    For the second part, we show that if $(V_s)_{s \in I} \in \mV_{I}^{(0)}$ then $s \mapsto \mL_{V_s} \, A$ is continuous with respect to $\lVert \cdot \rVert _{\nu, x}$ for all $\nu \in \N_0$ and $x \in \Z^d$ and that if $(V_s)_{s \in I} \in \mV_{I}^{(1)}$ then $s \mapsto \mL_{V_s} \, A$ is differentiable with respect to $\lVert \cdot \rVert _{\nu, x}$ for all $\nu \in \N_0$ and $x \in \Z^d$ with $ \tfrac{\d}{\d s} (\mL_{\Phi_s}\, A) = \mL_{\Dot{\Phi}_s} \, A$. The full statement then follows inductively. 
    
    So let $(V_s)_{s \in I} \in \mV_I^{(0)}$. Using Lemma \ref{lem: sum representation of generator} and the gauge invariance of $A$, we can write
    \begin{align*}
        \mL_{V_{t + h}} \, A - \mL_{V_t} \, A = \sum_{y \in \Z^d} [(v_{t+h} (y) - v_{t + h} (x) - v_t(y) + v_t(x)) \, n_y, \, A] \, .
    \end{align*}
    Now we observe that by Lemma \ref{lem:commutator bound}
    \begin{align*}
        \lVert [(v_t (y) - v_t (x)) \, n_y, \, A] \rVert_{\nu, x} &\leq 4^{\nu + d + 5} \, \frac{\lVert (v_t(y) - v_t(x)) \, n_y \rVert_{\nu + d + 2 , y} \, \lVert A \rVert_{\nu + d + 2, x}}{(1 + \lVert x - y \rVert)^{d + 2}}
        \\
        & \leq 4^{\nu + d + 5} \, \sup_{t \in I} \, C_{v_t} \, \frac{\lVert n_0 \rVert \, \lVert A \rVert_{\nu + d + 2, x}}{(1 + \lVert x - y \rVert)^{d + 1}} \, .
    \end{align*}
    This expression is independent of $t \in I$ and summable. Hence, the dominated convergence theorem applies and we can exchange the limit $h \to 0$ and the sum over $y \in \Z^d$. That is 
    \begin{align*}
        \lim_{h \to 0} \, \lVert \mL_{V_{t + h}} \, A - \mL_{V_t} \, A \rVert_{\nu, x} &\leq \sum_{y \in \Z^d} \lim_{h \to 0} \, \lVert [(v_{t+h} (y) - v_{t + h} (x) - v_t(y) + v_t(x)) \, n_y, \, A] \rVert_{\nu, x}
        \\
        & \leq 2 \sum_{y \in \Z^d} \lim_{h \to 0} \, (\lvert v_{t + h}(y) - v_t(y) \rvert + \lvert v_{t + h}(x) - v_t(x)\rvert) \, \lVert n_y \rVert_{\nu, x} \, \lVert A \rVert_{\nu, x} \, ,
    \end{align*}
    which vanishes, since the function $t \mapsto v_t(x)$ is continuous for every $x \in \Z^d$. Therefore we get the continuity of the map $s \mapsto \mL_{V_s} \, A$.
    
    Now let $(V_s)_{s \in I} \in \mV_I^{(1)}$. Since the map $s \mapsto v_s(x)$ is continuously differentiable for every $x \in \Z^d$, we can write
    \begin{align*}
        v_{s + h} (x) - v_s(x) = \int_{s}^{s + h} \d t \, \Dot{v}_t(x) \, .
    \end{align*}
    Thus as before we find
    \begin{align*}
        \left\| \frac{\mL_{V_{s + h}} \, A - \mL_{V_s} \, A}{h} - \mL_{\Dot{V}_s} \, A \right\|_{\nu, x} \leq 2 \sum_{y \in \Z^d} 4^{\nu + d + 5} \, \sup_{t \in I} C_{\Dot{v}_t}  \, \frac{\lVert n_0 \rVert \, \lVert A \rVert_{\nu + d + 2, x}}{(1 + \lVert x - y \rVert)^{d + 1}} \, .
    \end{align*}
    This allows us, again by dominated convergence, to pull the limit $h \to 0$ into the sum. This results in the differentiability of the map $s \mapsto \mL_{V_s} \, A$ and $\frac{\d}{\d s} (\mL_{V_s} \, A) = \mL_{\Dot{V}_s } \, A$.
\end{proof}

\medskip
\textbf{Acknowledgments}. 
This work was supported by the Deutsche Forschungsgemeinschaft (DFG, German Research Foundation) through TRR 352 (470903074) and FOR 5413 (465199066).

\printbibliography

\end{document}